\documentclass[conference]{IEEEtran}
\ifCLASSINFOpdf
  \usepackage[pdftex]{graphicx}
  \DeclareGraphicsExtensions{.pdf,.jpeg,.png}
\else
\fi
\usepackage{listings}

\usepackage{algorithmicx}
\usepackage{algpseudocode}
\usepackage{algorithm}

\usepackage{xcolor}

\usepackage[acronym]{glossaries}

\newacronym{hpc}{HPC}{High-Performance Computing}
\newacronym{cfd}{CFD}{Computational Fluid Dynamics}
\newacronym{gpu}{GPU}{Graphical Processing Unit}
\newacronym{fpga}{FPGA}{Field-Programmable Gate Array}
\newacronym{opm}{OPM}{Open Porous Media}
\newacronym{hbm}{HBM}{High-Bandwidth Memory}
\newacronym{hls}{HLS}{High-Level Synthesis}
\newacronym{uram}{URAM}{Ultra RAM}
\newacronym{pde}{PDE}{Partial Differential Equation}
\newacronym{spmv}{SpMV}{Sparse Matrix Vector Multiplication}
\newacronym{csr}{CSR}{Compressed Sparse Row}
\newacronym{rtl}{RTL}{Register-Transfer Level}
\newacronym{cg}{CG}{Conjugate Gradient}

\usepackage{multirow}
\usepackage{hhline}

\begin{document}
\title{Hardware Acceleration of HPC Computational Flow Dynamics using HBM-enabled FPGAs}

\author{
\IEEEauthorblockN{Tom Hogervorst\IEEEauthorrefmark{3}, Tong Dong Qiu\IEEEauthorrefmark{1}, Giacomo Marchiori\IEEEauthorrefmark{1}, Alf Birger\IEEEauthorrefmark{2}, Markus Blatt\IEEEauthorrefmark{4}, Razvan Nane\IEEEauthorrefmark{1}\IEEEauthorrefmark{3}}

\IEEEauthorblockA{\IEEEauthorrefmark{1}Big Data Accelerate B.V., Delft, The Netherlands}
\IEEEauthorblockA{\IEEEauthorrefmark{2}Equinor S.A., Bergen, Norway}
\IEEEauthorblockA{\IEEEauthorrefmark{4}OPM-OP AS, Oslo, Norway}
\IEEEauthorblockA{\IEEEauthorrefmark{3}Delft University of Technology, Delft, The Netherlands\\
r.nane@tudelft.nl}

}
% \and
% \IEEEauthorblockN{Homer Simpson}
% \IEEEauthorblockA{Twentieth Century Fox\\
% Email: homer@thesimpsons.com}
% \and
% \IEEEauthorblockN{James Kirk\\ and Montgomery Scott}
% \IEEEauthorblockA{Starfleet Academy\\
% San Francisco, California 96678--2391\\
% Telephone: (800) 555--1212\\


%Telephone: (800) 555--1212, Fax: (888) 555--1212}


% declaration of the new block
\algblock{ParWhile}{EndParWhile}
% customising the new block
\algnewcommand\algorithmicparwhile{\textbf{while}}
\algnewcommand\algorithmicpardo{\textbf{pardo}}
\algnewcommand\algorithmicendparwhile{\textbf{end\ while}}
\algrenewtext{ParWhile}[1]{\algorithmicparwhile\ #1\ \algorithmicpardo}
\algrenewtext{EndParWhile}{\algorithmicendparwhile}
% declaration of the new block
\algblock{ParFor}{EndParFor}
% customising the new block
\algnewcommand\algorithmicparfor{\textbf{for}}
\algnewcommand\algorithmicendparfor{\textbf{end\ for}}
\algrenewtext{ParFor}[1]{\algorithmicparfor\ #1\ \algorithmicpardo}
\algrenewtext{EndParFor}{\algorithmicendparfor}

\maketitle


\begin{abstract}
Scientific computing is at the core of many \gls{hpc} applications, including computational flow dynamics. Because of the uttermost importance to simulate increasingly larger computational models, hardware acceleration is receiving increased attention due to its potential to maximize the performance of scientific computing. A \gls{fpga} is a reconfigurable hardware accelerator that is fully customizable in terms of computational resources and memory storage requirements of an application during its lifetime. Therefore, it is an ideal candidate to accelerate scientific computing applications because of the possibility to fully customize the memory hierarchy important in irregular applications such as iterative linear solvers found in scientific libraries. In this paper, we study the potential of using \gls{fpga} in \gls{hpc} because of the rapid advances in reconfigurable hardware, such as the increase in on-chip memory size, increasing number of logic cells, and the integration of High-Bandwidth Memories on board. To perform this study, we first propose a novel ILU0 preconditioner tightly integrated with a BiCGStab solver kernel designed using a mixture of \gls{hls} and \gls{rtl} hand-coded design. Second, we integrate the developed preconditioned iterative solver in \textit{Flow} from the \gls{opm} project, a state-of-the-art open-source reservoir simulator. Finally, we perform a thorough evaluation of the FPGA solver kernel in both standalone mode and integrated into the reservoir simulator that includes all the on-chip URAM and BRAM, on-board \gls{hbm}, and off-chip CPU memory data transfers required in a complex simulator software such as \gls{opm}'s \textit{Flow}. We evaluate the performance on the Norne field, a real-world case reservoir model using a grid with more than $10^5$ cells and using 3 unknowns per cell, and we find that the \gls{fpga} is on par with the CPU execution and 3 times slower than the GPU implementation when comparing only the kernel executions.

\end{abstract}


% \fi
\IEEEpeerreviewmaketitle

\section{Introduction}

The advent of data proliferation and the slowing down of Moore's law is driving many application designers to implement an algorithm's computationally demanding functionality using a hardware accelerator. While \glspl{gpu} benefit from a larger user base because of their massive amount of parallel cores, on board \glspl{hbm}, and ease of programming, \glspl{fpga} are gaining recently a lot of attention because of their application-specific customizability. Furthermore, the recent integration of \glspl{hbm} in \glspl{fpga}-based systems coupled with the maturing of \gls{hls} programming tools \cite{surveyhls}, have touted the \gls{fpga} as a viable competitor for \glspl{gpu} in emerging machine learning-related domains, but also in \gls{hpc} applications where \glspl{gpu} were traditionally the sole candidate. The acquisition of Altera by Intel in 2016 \cite{intelaltera} and the recent announcement of AMD acquiring Xilinx \cite{amdxilinx}, two of the biggest \gls{fpga} providers, can be considered an indication of this trend.  

In this context, it is highly relevant to reassess the potential of \glspl{fpga} for \gls{hpc}, when we consider the addition of \gls{hbm} on board, the increase of on-chip memory (e.g., \gls{uram} \cite{uram}), and the continuing increase of  logic cell resources enabling one to implement complex functionality. To perform this evaluation, we look at an ubiquitous scientific computing field used in many \gls{hpc} application domains. Namely, the solution of systems of non-linear partial differential equations in time and space. These are typically not possible to solve exactly through analytical methods. Hence, the equations are discretized in time and over a domain space using finite difference, finite volume or finite element methods. Finally, the derived system of equations is solved, often iteratively using a combination of preconditioners and iterative linear solvers available in one of the many scientific solver libraries available, such as DUNE \cite{dunepaper2020}. PETSc \cite{petsclibrary} or others. Due to the complexity of implementing linear solvers on \gls{fpga}, there are only a few previous works that studied the acceleration of them on \gls{fpga}. However, many assumptions were made to simplify the design and implementation, such as either focusing solely on single precision floating point support, or using simple solvers without preconditioners, and all ignored the integration into the whole scientific computing pipeline. However, to realistically assess the potential of \glspl{fpga}, it is important to eliminate the aforementioned restrictions. Consequently, in this work we develop a complete solver with a preconditioner on \gls{fpga} with support for double precision, which is utterly important in a real-world setting because it enables the convergence to a solution. Furthermore, we integrate our work in a state-of-the-art reservoir simulator, \textit{OPM Flow} \cite{opmpaper}, and validate our work by running with a real-world use case containing a large number of grid cells ($10^6$). 

To accomplish our study goals, we develop a novel BiCGStab \cite{bicgstab} solver implementation with ILU0 \cite{ilu0} as preconditioner on \gls{fpga}. This combination of preconditioner and linear solver was chosen to get an apples-to-apples comparison with the CPU performance on the Norne field data set, as this combination was best performing in that case. To maximize the performance, we use a mixed programming model in which we combine \gls{hls} with manually written \gls{rtl}. Please note that there might be other solvers more efficient for a parallel architecture such as the \gls{fpga} (and \gls{gpu}) than an ILU0 preconditioner. However, this is left for future work because it was most important to make certain the results are close to the ones obtained by the already validated software only simulation. Therefore, we selected the same combination of preconditioner and solver as those used by default in \textit{OPM Flow} and designed the solver to support double precision because high accuracy is required, otherwise the selected real world case would not converge. Furthermore, integration is key to benchmark the performance in the overall system. Previous work has only benchmarked the solvers in isolation, whereas we consider the whole system for benchmarking, where including the memory transfers between off-chip and off-board is crucial. Finally, we make our work available in the \gls{opm} git repository \cite{opmrepo}.

The contributions of this paper are summarized as follows:

\begin{itemize}
    \item We propose a novel \gls{csr}-based encoding to optimize the \gls{spmv} on \gls{fpga}, which is the core functionality required to implement iterative solvers.
    \item We develop a hardware model to predict and guide the design of the solver, helping us to understand trade-off between area and performance when scaling the resources. 
    \item We design a novel ILU0-BiCGStab preconditioned solver on \gls{hbm}-enabled \gls{fpga} using a mixed programming model of \gls{hls} and \gls{rtl}.
    \item We integrate the proposed solver in \textit{OPM Flow}, both for \gls{fpga} and \gls{gpu}. Furthermore, we make all sources available in the \gls{opm} git repository. 
    \item We provide extensive evaluation results for both standalone solver kernel benchmarking as well as complete reservoir simulation execution on a real world use case running on three different types of systems, CPU, \gls{fpga}, and \gls{gpu}. 
\end{itemize}

In the next section \ref{sec:background}, we introduce the background required to understand the paper, while in section \ref{sec:relres} we highlight related works. We then start a bottom-up explanation approach in which we first describe the \gls{spmv} kernel in section \ref{sec:spmv}, then the solver in section \ref{sec:solver}, and discuss the performance hardware model used to guide the design in section \ref{sec:perfmodel}. Finally, in section \ref{sec:results}, we present the experimental results.

\section{Background}
\label{sec:background}

\subsection{Sparse Matrix Iterative Solvers} \label{background: solvers}

The central problem in this paper is to solve $A * \vec{x} = \vec{b}$ for vector $\vec{x}$, where A is a known sparse matrix and $\vec{b}$ is a known dense vector. There are different methods for solving such a problem, which can be grouped into direct solvers and iterative solvers. Direct solvers, as the name implies, use linear algebra techniques to solve the problem for x directly and precisely. However, direct methods are not suitable to solve large sparse linear systems, as it is usually the case in a real-world \gls{hpc} application, because they will fill-in the zero values of the sparse matrix in order to solve the problem, resulting in a very high computational cost and memory usage. Iterative solvers are a heuristic alternative to direct solvers: they start from an initial guess of the solution, and then iteratively improve this guess, until it is deemed close enough to the actual solution. In this paper, we use the BiCGStab (bi-conjugate gradient stabilized) solver with the ILU0 (incomplete LU factorization with no fill-in) as preconditioner because this combination is the most robust and the default software configuration used in \textit{OPM Flow}. Consequently, besides the goal of studying the acceleration potential of FPGAs, another implicit goal required by the reservoir engineers using the simulator was to obtain simulation results close to the software counterpart. Consequently, although it might not be the best choice for acceleration on a parallel architecture, fulfilling the conformance criteria led us to use the same solver used in the software stack.

Algorithm \ref{alg:BiCGSTAB} shows the pseudo-code of the BiCGStab solver. This code makes use of several other functions: 
\begin{itemize}
\item An \gls{spmv} function, which performs a Sparse Matrix-Vector (SpMV) multiplication
\item A dot function, which calculates the dot product (or inner product) of two vectors
\item An axpy function, which scales its first input vector by a scalar value, and adds it to its second input vector (i.e. $axpy(\alpha, \vec{x}, \vec{y})$ performs: $\alpha * \vec{x} + \vec{y}$).
\item A preconditioner function. This is an optional part of the solver, and does not have a set behavior. For more information about preconditioners, see section \ref{sec:preconditioner}.
\end{itemize}
The result of the solver are two vectors: $\vec{x}$, which is the estimation of the solution of $A * \vec{x_{actual}} = \vec{b}$, and $\vec{r} = \vec{b} - A \vec{x}$, a residual vector, which is an indication of how close the result $\vec{x}$ vector is to the exact solution $\vec{x_{actual}}$. It is worth noting in Algorithm \ref{alg:BiCGSTAB} how the exit condition of the solver is set, or in other words, how the algorithm determines when to stop iterating because the results are sufficiently close to the actual result. The solver stops iterating when the norm of the residual is lower than a certain \textit{convergence} threshold. There are two ways how this convergence threshold can be chosen: either it is a constant value, in which case we speak of an absolute exit condition, or it is a value set by the norm of the initial residual by a \textit{desired improvement} factor, in which case we speak of a relative exit condition. An absolute exit condition can be useful when one needs to be certain that the result of a solver is at least within a certain numerical distance from the actual result. However, when solving for multiple different matrices from different domains, a matrix with lower values all across the board might start off at a lower error and also reach the absolute exit condition more quickly. In such a case, a relative exit condition leads to a more fair comparisons between solves, which is why a relative exit condition is used in this paper.

Ignoring for now the preconditioner, what remains in the BiCGSTAB solver are a series of vector operations and \glspl{spmv}, as well as a few scalar operations. From profiling the solver, it is clear that the majority of its computational time is spent on performing \gls{spmv} and \gls{spmv} like computations. Therefore, the first goal to design a performant solver kernel would be to perform the \gls{spmv} efficiently. 

\begin{algorithm}                      % enter the algorithm environment
\caption{Pseudo-code BiCGSTAB solver}          % give the algorithm a caption
\label{alg:BiCGSTAB}              % and a label for \ref{} commands later in the document
\mbox{\textbf{Input}: $\vec{b}$, $\vec{x}$, $A$ \Comment{$\vec{b}$ and $\vec{x}$ are $N$ element arrays, A is an }}\\
\mbox{NxN matrix}\\
\mbox {\textbf{Output}:  $\vec{r}$, $\vec{x}$ \Comment{Both are $N$ element arrays}}\\
\begin{algorithmic}[1]                    % enter the algorithmic environment
    \State $\vec{r} = \vec{b} - spmv(A, \vec{x})$
    \State $\vec{rt} = \vec{r}$
    \State $\rho = 0$
    \State $\vec{p} = \vec{0}$
    \State $norm = sqrt(dot(\vec{r}, \vec{r}))$
    \State $conv\_{threshold} = norm / desired\_improvement$
    \While{$norm > conv\_{treshhold}$}
    \State $\rho_{new} = dot(rt, r)$
    \State $\beta = (\rho * \alpha) / (\rho_{new} * \omega)$
    \State $\vec{p} = axpy(\beta, \vec{r}, axpy(\omega, \vec{p}, \vec{v}))$
    \State $\vec{y} = preconditioner(\vec{p})$
    \State $\vec{v} = spmv(A, \vec{p})$
    \State $\alpha = \rho_{new} / dot(\vec{rt}, \vec{v})$
    \State $\rho = \rho_{new}$
    \State $\vec{x} = axpy(\alpha, \vec{x}, \vec{y})$
    \State $\vec{r} = axpy(-\alpha, \vec{r}, \vec{v})$
    \State $norm = sqrt(dot(\vec{r}, \vec{r}))$
    \If{$norm \leq conv\_{threshold}$}
    \State $break$
    \EndIf
    \State $\vec{y} = preconditioner(\vec{r})$
    \State $\vec{t} = spmv(A, \vec{p})$
    \State $\omega = dot(\vec{r}, \vec{t}) / dot(\vec{t}, \vec{t})$
    \State $\vec{x} = axpy(\omega, \vec{x}, \vec{y})$
    \State $\vec{r} = axpy(-\omega, \vec{r}, \vec{t})$
    \State $norm = sqrt(dot(\vec{r}, \vec{r}))$
    \EndWhile
\end{algorithmic}
\end{algorithm}

\subsection{Sparse Matrix-Vector Multiplication}
\label{sec:spmv}

The \gls{spmv} is a widely used, rather elementary operation in which a sparse matrix is multiplied by a dense vector. How this operation will be performed algorithmically depends on how the matrix is stored. A sparse matrix consist of many more zeroes than non-zero values, which means that storing all of them is not efficient. There are a variety of different formats in which only the non-zero values of a sparse matrix can be stored. Among these formats, Compressed Sparse Column (CSC) and Compressed Sparse Row (CSR) are two of the most commonly used, with both storing matrices efficiently without putting any restrictions on the content of the matrix, whereas other format may require the matrix to have certain sparsity pattern features such as symmetry, values being repeated multiple times, or store additional non-zeroes, e.g. sliced ELLPACK \cite{slicedellpack}, in order to be efficient. \gls{csr} and CSC do not have such restrictions. In this paper, we focus only on the \gls{csr} representation, as the column-major CSC format would be more difficult to use during the application of the ILU0 preconditioner because ILU0 goes through the matrix row-by-row (see section \ref{sec:preconditioner}), which would be difficult to implement in the column-major CSC format.

The \gls{csr} format consists of three arrays: an array with all non-zero values of the matrix stored row-major, an array of the column indices of those non-zero values, and an array of row pointers, which contains an index for each row to identify which non-zero value is the first value of that row, plus a last additional entry containing the total number of nonzeroes stored.
The pseudo-code showing how the SpMV is performed on a matrix stored in the CSR format is given in algorithm \ref{alg:SpMV}. It is worth noting that there is a high degree of parallelism available in this SpMV. The simple explication is that the calculations for all rows can be done in parallel because the result of one row is never used during the calculation of another row. Nevertheless, this is true only in theory because the irregular memory accesses in the x vector, which depend on the random column index representing the location of the non-zero value in the system matrix, restrict in practice the amount of parallelism that can be exploited efficiently. The main bottleneck is the requirement to replicate the x vector to enable multiple read units to feed independent row processing units. This replication is impractical for realistic large use cases for two main reasons: 1) limited on-chip (i.e., cache) memory and 2) significant preprocessing computations required to align the read data feed into the multiple processing units, which is needed to exploit the maximum available bandwidth in state-of-the-art accelerators with \gls{hbm}.

\begin{algorithm}                      % enter the algorithm environment
\caption{Pseudo-code of a CSR SpMV}          % give the algorithm a caption
\label{alg:SpMV}              % and a label for \ref{} commands later in the document
\mbox{\textbf{Input}: $nnzs$, $colInds$, $rowPointers$, $x$ } \\
\mbox{\Comment{$nnz$ is an $M$ element array of floating point values}}\\
\mbox{\Comment{$colInds$ is an M element array of integers}}
\mbox{\Comment{$rowPointers$ is an N+1 element array of integers}}\\
\mbox{\Comment{$x$ is an N element array of floating point values}}\\
\mbox {\textbf{Output}: $y$ \Comment{$y$ is an N element array of floating point values}}\\
\begin{algorithmic}[1]                    % enter the algorithmic environment
\For{$0 \leq i < N$}
\State $y[i] = 0$
\For{$rowPointers[i] \leq v < rowPointers[i+1]$}
\State{$y[i] = y[i] + nnzs[v] * x[colInds[v]]$}
\EndFor
\EndFor
\end{algorithmic}
\end{algorithm}

\subsection{The ILU0 Preconditioner}
\label{sec:preconditioner}

How hard a series of linear equations is to solve, depends on various different aspects of those equations, or the matrix that represents them. In other words, how many non-zeroes the matrix has, where they are located in the matrix, and their values in relation to one another, among other things, all play a role. The \textit{condition number} $\kappa(A)$ of a matrix $A$, which is the ratio of the biggest eigenvalue of the matrix divided by the smallest one, is a measure of how relative changes in the right hand side $b$ of a linear equation $A*\vec{x} = \vec{b}$ result in relative changes of the solution $\vec{x}$. In addition it a measure of how changes in $A$ affect the solution and how numerical errors affect the solution. Often, convergence of iterative methods is worse for matrices with high condition numbers. Therefore, while iterative solvers such as BiCGStab are a powerful method for solving sparse matrix problems, they might still take many iterations before they approach the result within the desired precision when solving matrices with a high condition number. 

One technique for reducing the amount of iterations needed for BiCGStab and other Krylov methods is utilizing a preconditioner. Generally, a preconditioner is a transformation or matrix $M$ that approximate $A$, but $M*\vec{x}=\vec{b}$ is less computationally expensive to solve than $A * \vec{x} = \vec{b}$. If $M$ is a matrix then the preconditioned method solves $M^{-1}A*\vec{x} = \vec{b}$. Note that if $M$ is a good approximation for $A$, then this system will have a much better condition number than the original one and hence the iterative will converge faster. A simple example of a preconditioner is the Jacobi method with $M=diag(A)$, the diagonal values of $A$. Many different types of preconditioners exist, from very simple ones that only very weakly approximate the problem's solution, to complex ones that almost solve the system by themselves. Unfortunately, how well a preconditioner performs depends not only on the type of preconditioner, but also on the matrix being solved. For example, the Jacobi preconditioner is fairly accurate if the original matrix has most of its non-zeroes clustered around the diagonal, and if the values on the diagonal have a higher magnitude than those off the diagonal, but does not work as well for matrices that do not have those characteristics. 

In this paper, we utilize the zero fill-in incomplete LU factorization (ILU0) as preconditioner, which falls somewhere in the middle: it approximates the system reasonably while not being too computationally intensive to apply. Using an ILU0 consists of two steps: first, before the solver is started, an incomplete LU decomposition is performed. To understand what this is, we must first introduce the LU decomposition. LU decomposition is a direct solving method that calculates an upper-triangular matrix U and a lower-triangular matrix L so that $A = L * U$. Such a decomposition can be done using Gaussian elimination, among other methods. Once such a decomposition has been found, $A*\vec{x} = \vec{b}$ can be solved for $\vec{x}$ by first solving $L*\vec{y} = \vec{b}$ for $\vec{y}$ and then solving $U*\vec{x} = \vec{y}$ for $\vec{x}$. Since L and U are triangular matrices, these two solves can be done using forward and backward substitution respectively. However, like we discussed in the previous section, direct solving method become very computationally intensive for sparse matrices because they start filling in the zero values of the matrix, and LU decomposition is no exception. Incomplete LU(N) decomposition decreases the high computational cost of LU decomposition for sparse matrices by reducing the amount of fill-in that happens, where N is an indicator of how much fill-in is done. The case were $N = 0$ (ILU0) means that for the L and U matrices the decomposition generates non-zero values only at the same indices where A had non-zero values.

Once the ILU0 decomposition is complete, the solver can start. The solver will then apply the ILU0 preconditioner in every iteration. Pseudo-code of the ILU0 application is shown in algorithm \ref{alg:ILU0_apply}. The ILU0 application consists of two parts, represented by the two for loops in lines 1 and 11: a forward substitution on the lower-triangular part of the LU matrix, and a backward substitution on the upper-triangular part of that matrix. Both loops are very similar in the operations they perform, with the main differences being that the forward substitution traverses the matrix top-to-bottom, while the backward substitution traverses the matrix bottom to top, and that the backward substitution concludes by dividing the result array by the values on the diagonal of the LU matrix. It is worth noting that while these loops ostensibly have the same basic structure as the SpMV loop, there is one major difference: the results of one row depend on rows before it, which means the operations on the rows can not be performed all in parallel. See section \ref{sec:graph_coloring} for more on how this issue can be circumvented somewhat by re-ordering the matrix.

\begin{algorithm}                      % enter the algorithm environment
\caption{Pseudo-code of the applying of the ILU0 preconditioner}          % give the algorithm a caption
\label{alg:ILU0_apply}              % and a label for \ref{} commands later in the document
\mbox{\textbf{Input}: LUmat, $x$ } \\
\mbox{\Comment{LUmat is an NxN sparse matrix in the CSR format}}\\
\mbox{\Comment{$x$ is an N element array of floating point values}}\\
\mbox {\textbf{Output}: $p$ \Comment{$p$ is an N element array of floating point values}}\\
\begin{algorithmic}[1]                    % enter the algorithmic environment
\For{$0 \leq i < N$}
\State $p[i] = x[i]$
\For{LURowPointers$[i] \leq v < $ LURowPointers$[i+1]$}
\State $pIndex =$ LUcolInds$[v]$
\If{$pIndex \geq i$}
\State $break$
\EndIf
\State{$p[i] = p[i] - $LUnnzs$[v] * p[pIndex]$}
\EndFor
\EndFor

\For{$N > i \geq 0$}
\State $diagIndex =$ LURowPointers$[i+1]$
\For{LURowPointers$[i+1] > v \geq $ LURowPointers$[i]$}
\State $pIndex =$ LUcolInds$[v]$
\If{$pIndex \leq i$}
\State $diagIndex = v$
\State $break$
\EndIf
\State $p[i] = p[i] - $LUnnzs$[v] * p[pIndex]$
\EndFor
\State $p[i] = p[i] / $LUnnzs$[diagIndex]$
\EndFor
\end{algorithmic}
\end{algorithm}

To quantify the effect of the ILU0 preconditioner in practice for our solver performance , we ran software implementations of the solver without preconditioner and with the Jacobi and ILU0 preconditioners on several benchmark matrices. The matrices were obtained from the SuiteSparse matrix collection \cite{suitesparse, suitesparsepaper}.

\begin{table}[!h]
\caption{cpu solver time in ms with $10^{-2}$ \& $10^{-6}$ exit condition}
\begin{center}
\label{tab:preconditioners}
\begin{tabular}{|l||c|c||c|c||c|c||}
\hline
Precond. & None & None & Jacobi & Jacobi & ILU0 & ILU0 \\ 
\hline
Matrix & $10^{-2}$ & $10^{-6}$ & $10^{-2}$ & $10^{-6}$ & $10^{-2}$ & $10^{-6}$ \\ 
\hline
Hummocky & 5.9 & 925 & \textbf{1.2} & 125.8 & 3.1 & \textbf{32.3} \\
bcsstk25 & 5.8 & 3396 & \textbf{3.7} & \textbf{34.9} & 18.9 & 4331\\
bodyy6 & 4.4 & 224 & \textbf{1.2} & 86.3 & 3.7 & \textbf{15.7} \\
wathen120 & 8.0 & 112 & \textbf{5.2} & 25.2 & 12.8 & \textbf{18.7} \\
gridgena & \textbf{2.0} & 403 & 5.9 & 643.3 & 18.7 & \textbf{324.7} \\
qa8fm & \textbf{12.7} & 114 & 13.1 & 79.7 & 44.8 & \textbf{60.3} \\
\hline
\end{tabular}
\end{center}
\end{table}
%\vspace*{-10mm}

Table \ref{tab:preconditioners} shows the results of these tests, which were obtained by running the preconditioned solver for a single-thread on a Intel i7-9700 CPU with 16 GB of RAM. Two exit conditions were used: the more precise exit condition requiring the absolute value of the residue being reduced by $10^{-6}$ of its initial value, and the less precise exit condition requiring a residue norm reduction of $10^{-2}$. Not shown in the table are the number of iterations in which the solvers converged, which in all matrices except for \textit{bcsstk25} were lower when ILU0 preconditioning was used over the other two preconditioning options. However, a lower iteration count does not directly result in a higher solver performance. The first reason for this is the time the pre-processing of the preconditioner takes, which accounts for between 60 \% and 90\% of the time taken by the lower-accuracy solver with ILU0 preconditioner. The Jacobi precoditioner has barely any pre-processing time, and is powerful enough to be preferred over no preconditioner for all but one matrix, and preferred over ILU0 for all matrices when the solver's precision is low. Only when the solver's precision was set higher, did the ILU0 preconditioner's power show: the number of iterations required to reach this accuracy was high enough that the pre-processing time no longer dominated the run time. For most matrices, the ILU0 preconditioner reduced the iteration count enough that the solve performance was better than that of the other preconditioner options tested. The bcsstk25 matrix being an exception, as its solve was only accelerated by the Jacobi preconditioner. We suspect the matrix' narrow banded structure and high values on the diagonal make it particularly well-estimated by the Jacobi preconditioner. The second reason for why a lower iteration counts doesn't translate to a higher solver performance directly is not captured in our software tests. That is, different preconditioners might be easier or harder to parallelize; therefore, making them more or less attractive to be accelerated on hardware respectively. The single vector calculation that is the applying of the Jacobi preconditioner is easy to parallelize, and therefore attractive for hardware acceleration. ILU0 application on the other hand is not that inherently parallel, but can be made more parallel when using graph coloring.

\subsection{Matrix Reordering}
\label{sec:graph_coloring}

As covered in section \ref{sec:preconditioner}, the ILU0 application function does not inherently have a lot of parallelism for a hardware accelerator to exploit. However, by reordering the matrix, some parallelism in this function can be obtained. The goal of such a reordering is to group rows together that do not depend on one another (i.e., that do not have non-zero values in columns that have the same index as the row indices of other rows in the group). During our work, we have focused on two ways of doing such a reordering: level scheduling and graph coloring. 

\subsubsection{Level Scheduling} 
Level scheduling (LS) is a way of re-ordering a matrix by moving the rows and the columns of the matrix. That is, the rows are grouped together in colors that can be operated on in parallel, and then the columns are reordered in the same way to keep the matrix consistent in the (re)order of the unknowns, which is required during ILU decomposition. LS does not only group independent rows together into 'levels', but that also keeps the order of dependencies of the original matrix intact \cite{saad2003LS}. In other words, if a certain row j that depends on row i comes after row i in the original order, then after level scheduling, row $j_{ls}$ (the row to which row j was moved during level scheduling) will still come after $i_{ls}$ for all possible values of i and j. A level-scheduling algorithm will go through the matrix and gather all rows that do not depend on any other rows into the first level. Then it will go through the matrix again, and group all rows that only depend on rows in level 0 into level 1. Then it will fill level 2 with rows only dependent on rows from levels 0 and 1, and so on until every row has been added to a level. 

\subsubsection{Graph Coloring}

Graph Coloring (GC) is a term that we use to describe the reordering of matrices into groups ('colors') of independent rows that do not require the maintaining the order of dependencies in a matrix. To be more specific, we use the Jones-Plassmann algorithm \cite{Jones1993GraphColoring} to do graph coloring. This algorithm assigns a random value to every row in the matrix, and then groups all rows that have a higher number than all not-yet-grouped rows it depends on into a single color, and keeps repeating this process until all rows have been assigned a color. In this work, graph coloring was chosen over the other method because it is parallelizable and it allows more control over the maximum amount of rows and columns in a color, which will alleviate some of the inherent hardware design limitations. The disadvantage of graph-coloring is that it is not guaranteed to find the minimum number of colors, and thus the highest amount of parallelism because of its random nature. 

\subsubsection{Reordering Scheme Selection}
Selecting one of the two matrix reordering methods depends on the answer to how important it is to obtain a solution that is as close as possible to the one obtained without reordering. If the answer is \textit{very important}, then level scheduling is the preferred reordering scheme. However, when the answer is \textit{not important}, because an iterative solver is already estimating the result instead of solving the system precisely, graph coloring can be used if the result is \textit{close enough} to the actual solution.

\subsection{Sparstitioning}

To increase the scalability of our design and allow it to  work on larger matrices, we will use sparstitioning \cite{Sigurbergsson2019SparstitionAP}. Sparstitioning is a method for splitting both the matrix and the vector of a matrix-vector operation into partitions, so that no unnecessary data is read during the running of the operation on each partition. However, we make two changes to the sparstitioning algorithm. Firstly, we modify how the partitions are chosen to conform with the matrix reordering scheme chosen. Concretely, every color/level will be one partition. Secondly, we modify how the vector partitions are stored. When sparstitioning is used for an SpMV, the vector that is operated on is stored in a partitioned format, which introduces some duplication of the vector data. However, storing the vector in this way is not convenient because the non-partitioned vector is needed for the vector operations of the solver. Therefore, instead of storing the vector partitions themselves, our design stores an array that denotes for every partition which indices of the vector are part of that partition. Consequently, only those values can be read from the vector when computations start on that partition.

Therefore, because of the matrix reordering and sparstitioning optimization required to design a \gls{hpc} iterative solver, we modify the standard CSR matrix storage format in two ways. Firstly, the number of non-zeroes, rows, and vector partition indices used in each partition of the matrix need to be known. These will be collectively referred to as the sizes of each color. Secondly, the vector partition indices themselves are needed, so that the solver knows which values of the matrix are accessed by each color. To account for these CSR modifications, we introduce our own format, CSRO, which we will formally introduce in section \ref{matformat}.

\section{Related works}
\label{sec:relres}

\subsection{Sparse Matrix-Vector Multiplication on FPGAs}
\label{sec:relresspmv}

As described in section \ref{background: solvers}, in this paper we focus on \gls{cg} solvers, with a focus on its general version (i.e., not only for symmetric matrices as CG is), BiCGStab. Previous research into accelerating (parts of) CG solvers using FPGA has focused until now mostly on accelerating the SpMV kernel for one good reason: The SpMV function has a wide range of uses in a broad variety of domains, and because of its irregular memory accesses, it is not efficiently performed by classical von-Neumann architectures like CPUs. This makes SpMV a perfect candidate to be accelerated using customizable dataflow architectures that can fully exploit application-specific configurable memory hierarchies. Furthermore, in the context of linear algebra iterative solvers, SpMV takes up most of time. Consequently, a plethora of previous work addressed the topic of accelerating SpMV on FPGAs. Therefore, we only list a few of the performant designs without trying to be exhaustive.

\begin {itemize}
\item The SpMV unit proposed by Zhuo and Prasanna \cite{zhuo2005sparse} achieves a performance of between 350 MFLOPS and 2.3 GFLOPS, depending on the number of processing units, the input matrix, the frequency, and the bandwidth. However, to realize their design, the authors use zero-padding and column-wise blocking of the matrix. These design choices make SpMV less efficient due to storage overhead that becomes critical for large matrices and very inefficient to use in combination with an ILU0 preconditioner.
\item Fowers et al. \cite{fowers2014high} implemented a SpMV unit that uses a novel sparse matrix encoding designed with the fact that the matrix will be divided over multiple processing units in mind, and a banked vector buffer that allows many memory accesses to the input vector at the same time. That implementation reaches a performance of up to 3.9 GFLOPS, but does not actually outperform GPU or CPU implementations for large matrices.
\item Dorrance et al. obtain a performance of up to 19.2 GFLOPS \cite{dorrance2014scalable} for their CSC based SpMV accelerator, which is as far as we could find the highest performance of any FPGA-based implementation of SpMV. This high performance was thanks to the high computational efficiency of the implementation paired with its 64 processing units and high bandwidth (36 GB/s). However, as can be seen in algorithm \ref{alg:ILU0_apply}, in the applying of the ILU0 preconditiner, the result of one row is needed to calculate the result of ones after it. So, unlike the SpMV, the apply\_ILU0 can not be done as easily with CSC as it can with CSR. In fact, the CSC is completely unsuitable to use in conjunction with ILU0, and as a result we will not use that storage format in this paper.
\end{itemize}

Please note that the scope of the paper is not to design the fastest SpMV unit possible, but also to integrate this in a larger and complex application that is a preconditioned linear solver. Consequently, although some of the design choices might be sub-optimal for the design of the \textit{fastest} SpMV (e.g., using a CSR like format instead of the more parallelizable CSC format), which is thus not a primary goal, we focused on the complete integration and looked for the best SpMV design options from a system-level view, and not SpMV in isolation as most of previous work considers.
Nevertheless, the SpMV designed and implemented in this work is on par with the best SpMV solution we could find in the literature when we consider normalized resources and bandwidth as comparison parameters. Our SpMV is able to achieve up to 4.5 and 8.5 GFLOPS in double and single precision, respectively.  

\subsection{Sparse Matrix CG Solvers on FPGAs}

Research into designing a complete CG solver on an FPGA has been done as well. However, probably due to the complexity of iterative solver algorithms coupled with a limited area, small on-chip memory, and no HBMs available on previous generation FPGAs, we could find only a handful of related works of solvers on reconfigurable hardware. 
\begin{itemize}
\item Morris et al. \cite{morris2006hybrid} implement a CG solver on a dual-FPGA using both a HLL-to-HDL compiler and custom floating point units written in VHDL. Despite the fact that using the HLL-to-HDL compiler is probably less optimized that writing VHDL directly, they obtain a speed-up over software of 30\%, showcasing the potential of FPGA in accelerating sparse CG solvers.
\item Wu et al. \cite{wu2013high} implemented a CG solver on an FPGA consisting of an efficient SpMV unit, a vector update unit, and a diagonal preconditioner. They obtain a speedup between 4 and 9 times over their software implementation.
\item Chow et al. \cite{chow2014efficient} introduce a novel way of statically scheduling accesses to the on-chip memory in a \gls{cg} solver on an \gls{fpga}. Despite the initial cost of performing this scheduling, an average speedup of 4.4 over a CPU solver, and of 3.6 over \gls{gpu} one is achieved. 
\end{itemize}

However, different from the work in this paper, all of this previous research focused on small scale matrices that are not enough for a HPC real-world application setting. Furthermore, another thing that all of these implementations share is that no or only a very simple preconditioner has been used. Preconditioners can improve the performance of a solver significantly, as they reduce the amount of iterations that the solver needs to run until it reaches its desired precision significantly. Despite costing additional time to perform each iteration, the ILU(0) preconditioner used in this work improves the performance of all kinds of CG solvers over cases without preconditioners or with the diagonal preconditioner \cite{topsakal2001evaluation}. Consequently, not using a preconditioner in an iterative solver for a HPC real-world application that requires very large and sparse matrices, is not an efficient approach. Finally, using the CG solver limits the applicability of the design to only problems that use symmetric matrices to describe the physical system.

\section{SpMV multiplication}
\label{sec:spmv_mult}

\subsection{Design Objective}
The main objective of our SpMV kernel on the FPGA is to ensure that the unit has a high computational efficiency, which is required to fully exploit the \gls{hbm} available on the state-of-the-art FPGA generations such as the Xilinx Alveo U280 \cite{u280} data center accelerator card. The remainder of this section highlights the design choices of the SpMV implementation for the \gls{hpc} domain when using the latest \gls{hbm}-enhanced \gls{fpga} accelerators.

\subsection{The Matrix Format}
\label{matformat}

As described in section \ref{sec:relresspmv}, we have chosen to base our kernel on the row-major CSR matrix storage format. Furthermore, CSR is also the default format used in \textit{Flow}, a real-world HPC software application that we will use in section \ref{sec:results} to validate our results and benchmark the performance. However, this format poses a challenge for any HPC FPGA-based design, namely because the CSR contains one integer per matrix row to describe how many values are each row. Since our design objective is to maximize throughput of the SpMV unit, we want to feed as many non-zero values and column indices into the SpMV unit as it can sustain on every clock cycle. Unfortunately, it is not straight-forward when using the CSR's \textit{rowPointers} to determine to which row each value belongs to. The number of rows a given number of non-zero values in a matrix belong to can vary wildly based on the matrix's sparsity pattern. Consequently, the number of rowPointers that need to be read every cycle also varies, and also depends on the values of the rowPointers of the cycle before it. For this reason, we introduce a novel sparse matrix storage format based on CSR to be used by our accelerator. 

\begin{figure}[!h]
    \centering
    \includegraphics[width=0.94\columnwidth]{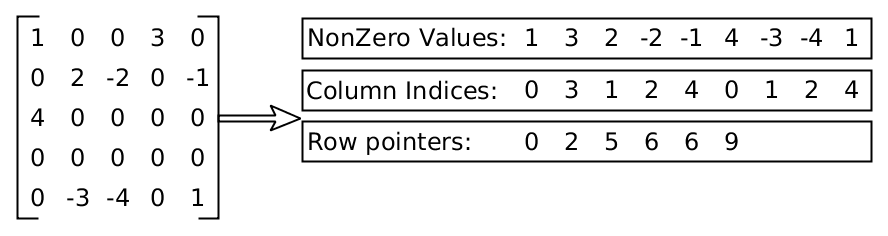}
    \caption{An Example Matrix Stored in the CSR Format.}
    \label{fig:CSR_example}
\end{figure}

\begin{figure}[!h]
    \centering
    \includegraphics[width=0.94\columnwidth]{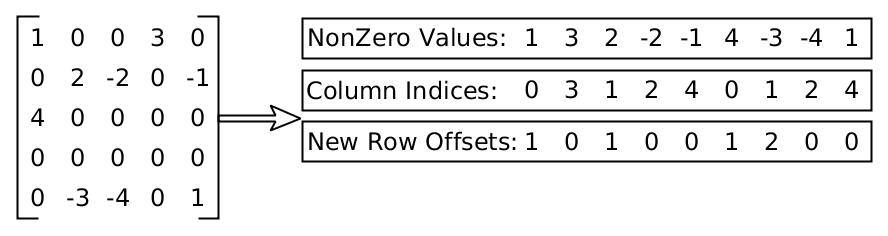}
    \caption{An Example Matrix Stored in the Novel CSRO Format.}
    \label{fig:CSRO_example}
\end{figure}

Similar to CSR, this new format contains the matrix's non-zero values and their column indices as two arrays, but instead of the rowIndices, it contains a new row offset for every nonzero value in the matrix. Because of the added offset array, this new format is tentatively called the \textit{Compressed Sparse Row-Offsets (CSRO)} format. A value in the newRowOffset array is 0 if the non-zero value at that same index is in the same row as the non-zero value before it. If a non-zero value is not in the same row as the value before it, then the rowOffset at the same index will be equal to 1 + the number of empty rows between that non-zero value's row and its predecessor's row. To help illustrate how a matrix is stored in the CSRO format, figure \ref{fig:CSRO_example} shows a matrix as it would be represented in the CSRO format. For comparison, figure \ref{fig:CSR_example} shows how that same matrix would be represented in the CSR format. Note that the NewRowOffset value of the first value in the CSRO format is 1. This is a convention adopted to show that the first row starts with the first value. 

\subsubsection{The Blocked CSR}

The OPM platform and its flow simulator do not use regular CSR, but rather blocked CSR, in which each non-zero value in the matrix is not one value, but a 3x3 matrix of values. This is done to group all relevant physical information about a single point of the reservoir that the matrix modelled together, but that is not relevant for this paper. What is relevant, is that the nine values in the 3x3 blocks are not easily used on an FPGA. Not only do these blocks often contain zeroes (about half of all values in the blocks of the NORNE testcase are zeroes), but the odd number of values in a block poses a problem when reading them over the ports of the FPGA or storing them in its internal memories, all of which are build to consist of an even number of values. Since the new matrix format that we introduce and the sparstitioning already require extensive reordering of the non-zero values of the matrix, we take this opportunity to also remove the blocking from the matrices in those steps. As a result, the FPGA solver kernel primarily works on non-blocked matrices. The only exception to this is the division at the end of the backwards substitution step in the apply\_ILU0. Here, a division by the value on the diagonal need to be done, and this value on the diagonal is a block. These diagonal blocks are the only blocks that could not be removed, and as a result the hardware kernel still uses those blocks as inputs.

\subsection{Design Overview}

\subsubsection{The SpMV Pipeline}

\begin{figure}[!h]
    \centering
    \includegraphics[width=0.94\columnwidth]{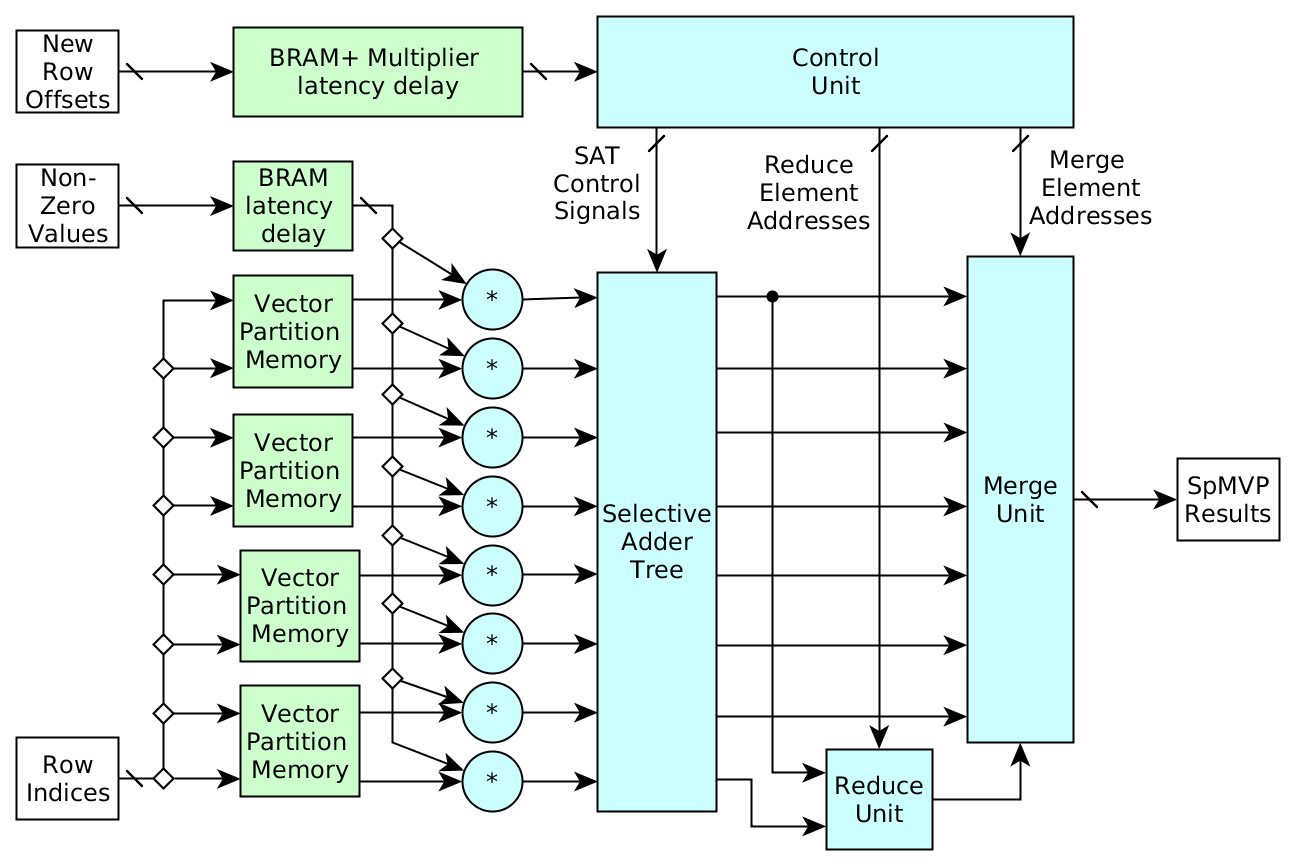}
    \caption{A High-Level Schematic of the SpMV Unit.}
    \label{fig:spmv_high_level}
\end{figure}

The upper limit of the throughput of our SpMV unit design is determined by the number of multipliers in the unit. Thus, to achieve a high computational efficiency, our design aims to stream values into those multipliers every cycle. Figure \ref{fig:spmv_high_level} shows how data is streamed into the multipliers: non-zero values and column indices are read from the memories that hold them, and the column indices are used as addresses at which to read elements from the input vector, while the non-zero values are delayed by the delay of the vector memory. In this way, the non-zero matrix values and the corresponding vector elements arrive in the multiplier at the same time.  There is one memory that holds the vector partition for every two multiplier units, because the BRAM blocks of an FPGA have at most two read ports. Which values of the vector are part of a certain partition is determined by the vector partition indices of that partition. Those values are calculated by the software during pre-processing. Each vector partition memory contains the entire vector partition, so the data is replicated and stored in each one. While not the most space-efficient solution, this does allow a high number of simultaneous memory accesses at different addresses, which is what the SpMV unit requires.

After the multiplications have finished, the SpMV unit needs to add together the multiplication results that belong to the same matrix row together to obtain the SpMV result of that row. For this, we designed a selective adder tree (SAT) and a reduce unit, which receive instructions about which values to add together from a control unit. This control unit in turn obtains this information from the new row offset data. The selective adder tree adds all the values in a single cycle that belong to the same row together, while the reduce unit adds together the SAT results from different clock cycles that belong to the same row. SAT results that already represent the final result of a row (i.e. if all the values of that row were in the same clock cycle) do not need to go into the reduce unit. Finally, a merge unit merges all of the results of the adder tree and reduce unit into a set number of output ports that will be written to a result memory.

\subsubsection{SpMV Top Level}

While the SpMV pipeline is the most important part of the SpMV unit, as it does the calculations, some additional units are needed to keep the pipeline running correctly. These units are as follows:
\begin{itemize}
    \item The external read unit handles all reads from outside of the FPGA chip. At the start of the SpMV run it reads the sizes of all colors, and uses these to read all matrix data that is needed during the SpMV.  
    \item For the SpMV unit to be able to use sparstitioning with vector partition indices, the vector values at those indices need to be able to be read quickly. To achieve this, our FPGA design contains a  single large URAM memory to store the multiplicant vector once. Due to the current size chosen for this memory in the design, up to 262144 values can be stored, meaning that a matrix may not have more than 262144 columns to be able to be multiplied by the kernel. The internal read unit coordinates reads from this memory. It receives the vector partition indices of the current color from the external read unit, reads the vector values at those indices, and sends those values into vector partition memories of the SpMV pipeline.
    \item The results of the SpMV pipeline can be out of order, as the time some result spend in the reduce and merge units are different than others. The write unit stores the results of the SpMV pipeline in a memory block that has been cyclically partitioned, so that it can write multiple results into it and at the same time still read from it. It also keeps track of up to which index all results have been received. Whenever it has received enough values to fill up a cacheline, it will queue up that line to be sent to the HBM memory. 
\end{itemize}

\begin{figure}[!h]
    \centering
    \includegraphics[width=0.9\columnwidth]{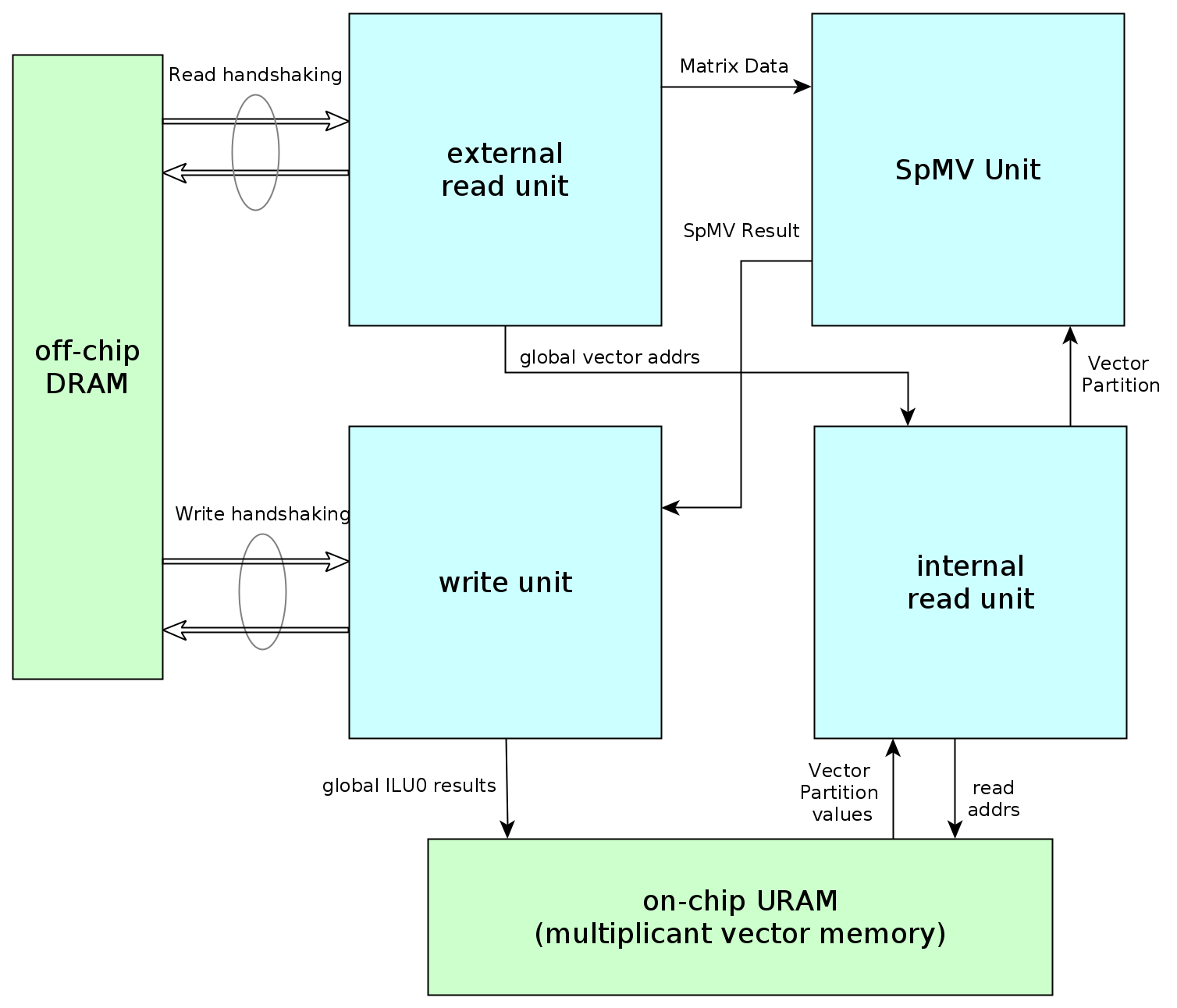}
    \caption{Schematic SpMV Top-level }
    \label{fig:spmv_schematic}
\end{figure}

An overview of the SpMV top-level unit is shown in figure \ref{fig:spmv_schematic}. Not shown in this figure is the control logic around it that instructs the units how much data to read or write and to/from where. To do this, the control logic keeps track of which color it is currently operating on and how many still remain.

\subsubsection{Non-partitioned SpMV}

One of the tasks the SpMV top-level unit must perform is the reading of the vector partition data that the sparstitioning adds to the matrix data. For a large part, this read operation can be done as look-ahead, meaning that while the SpMV pipeline operates on one color, the vector partition data of the next color can already be read from the URAM. This data is then stored in a BRAM block in the internal read unit. However, after the SpMV pipeline completes, this vector partition data needs to be transferred to the vector partition memories of the SpMV pipeline, and the vector partition indices for the next color need to be read from the off-chip memory. This is overhead added by our choice to use sparstitioning. To determine how much this overhead slows down the overall performance, we added the option to the design to not use sparstitioning. This version of the design has as downside that it has a more limited maximum matrix size that it can operate on, as the complete multiplicand vector need to be stored multiple times in the vector memories of the SpMV pipeline. The advantage is that these additional steps of reading vector partition indices and transferring vector partition data no longer need to occur.

\begin{figure}[!h]
    \centering
    \includegraphics[width=0.9\columnwidth]{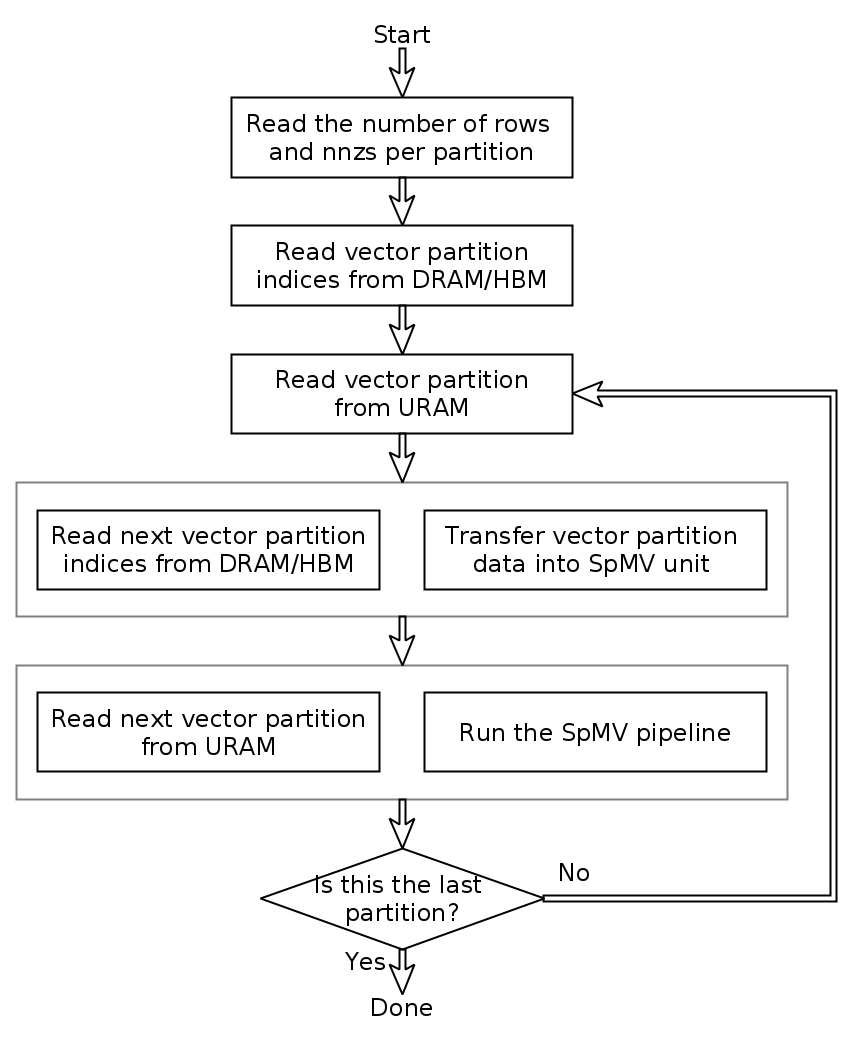}
    \caption{Overview of the Tasks in the Sparstitioned SpMV Unit.}
    \label{fig:spmv_sparstitioned}
\end{figure}

\begin{figure}[!h]
    \centering
    \includegraphics[width=0.4\columnwidth]{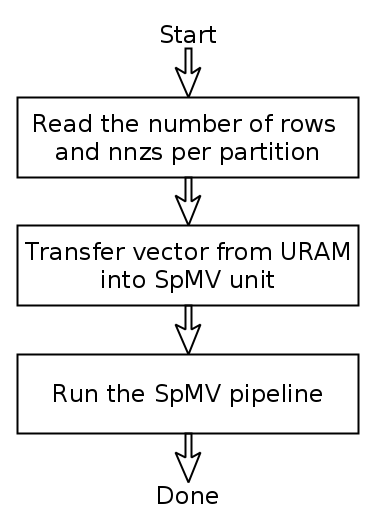}
    \caption{Overview of the Tasks the Non-Sparstitioned SpMV Unit.}
    \label{fig:spmv_non_sparstitioned}
\end{figure}

To visualize the difference in way of operating between the sparstitioned and the non-sparstitioned SpMV unit, figure \ref{fig:spmv_sparstitioned} shows the steps the sparstitioned SpMV unit takes, while figure \ref{fig:spmv_non_sparstitioned} shows those the non-sparstitioned SpMV unit takes. Tasks horizontally next to one another in these figures are performed at the same time. Obviously, the flow in figure \ref{fig:spmv_non_sparstitioned} is much simpler. Also note how in figure \ref{fig:spmv_sparstitioned} the reading of the next vector partition (and the reading of the indices to read from) are already done while the SpMV unit is still working on another partition. This hides the time these reads would normally take, but also necessitates an additional memory to read the next vector partition into, and a step in which the data is transferred from this memory into the SpMV unit.

\subsection{GPU SpMV kernels}
For the GPU, kernels in CUDA and OpenCL are tested. Only doubles are used for nonzero values.
NVIDIA's cuSPARSE \cite{cusparse} features a blocked SpMV function called cusparseDbsrmv().
For OpenCL, algorithm 1 from \cite{bcsr_spmv_paper} is used.
In this algorithm, a warp of 32 threads is assigned to a single block row, and threads are assigned to elements in the block row.
To cover the entire block row, a warp iterates, handling $floor(32/bs^2)$ blocks at a time, where bs is the blocksize.
A warp only operates on complete blocks, so some threads might be always idle.
For 3x3 blocks, a warp can cover 3 blocks with 27 threads, and has 5 inactive threads.
During initialization, a thread calculates the index of the block row, how many blocks the row has, and which block it is supposed to operate on.
It also calculates its position within a block (row and column).
Since a warp only covers complete blocks, a thread's assigned position in a block never changes.
During iterating, it multiplies the target element from A with the value from x, adding the product to a running total. Once the whole block row is complete, threads with the same vertical position (row) in the blocks reduce the values in their local registers and write the reduced output to global memory. Reduction can be done with shuffling or via shared memory.

\section{Sparse Matrix solver design}
\label{sec:solver}

To expand the SpMV kernel into a kernel that can perform preconditioned BiCGStab, units to perform both the preconditioning and the vector operations need to be added. In this section, the design of those units is discussed, followed by a description of the complete design that links all units and performs the complete preconditioned solve.

\subsection{Applying the ILU0 Preconditioner}

As shown in algorithms \ref{alg:SpMV} and \ref{alg:ILU0_apply}, the basic structure of the SpMV and the forward and backward substitutions that the ILU0 application performs are similar. This, combined with the fact that the SpMV and ILU0 application are never active at the same time, since the result of one is needed to start the other, makes re-using the hardware of the SpMV unit for the ILU0 apply a desired feature to save resources. Unlike the SpMV, which goes through the matrix it operates on line-by-line, the ILU0 application goes through its LU matrix in a no-sequential order. First, it goes through all of the values below the diagonal top-to-bottom, and then it goes through all values above and including the diagonal bottom-to-top (last row first). This memory access pattern is not efficient to use on the FPGA because it goes through the LU matrix skipping the second half of the values of each row during the forward substitution, since it only uses the lower triangular matrix there. Then, it goes back up through the matrix only using the second half of the values, since it only needs the upper triangular matrix during the backward substitution. Consequently, it needs to read all LU matrix data twice, ignoring about half of it each time.  Which data it needs to use and which it doesn't depends on the current row number and the column index. This memory access pattern is not specifically inefficient for the FPGA, but just in general, so in our design the lower-triangular matrix L, the upper triangular matrix U, and the diagonal values are split from the LU matrix, and used as three separate data structures. First, L is read to perform the forward substitution, and then U and the diag$\_$vals are read to perform the backward substitution. Since the latter step happens bottom-to-top, both the U matrix and the diag$\_$vals are stored in reverse order.

The main difference between the substitutions and the SpMV is that the substitutions modify the vector that they are operating on, while the SpMV doesn't. All rows that are grouped into the same color by graph coloring can be in the pipeline at the same time, but between the colors, the results of the previous color need to be integrated into the vector that is currently being operated on. To achieve this, the results of the ILU0 application are not only written to the URAM vector memory, but also forwarded directly into the vector partition memories so that they can be used during the next color, without requiring the unit to wait with reading its vector partition until the previous color finishes its computation.

\begin{figure}[!h]
    \centering
    \includegraphics[width=0.9\columnwidth]{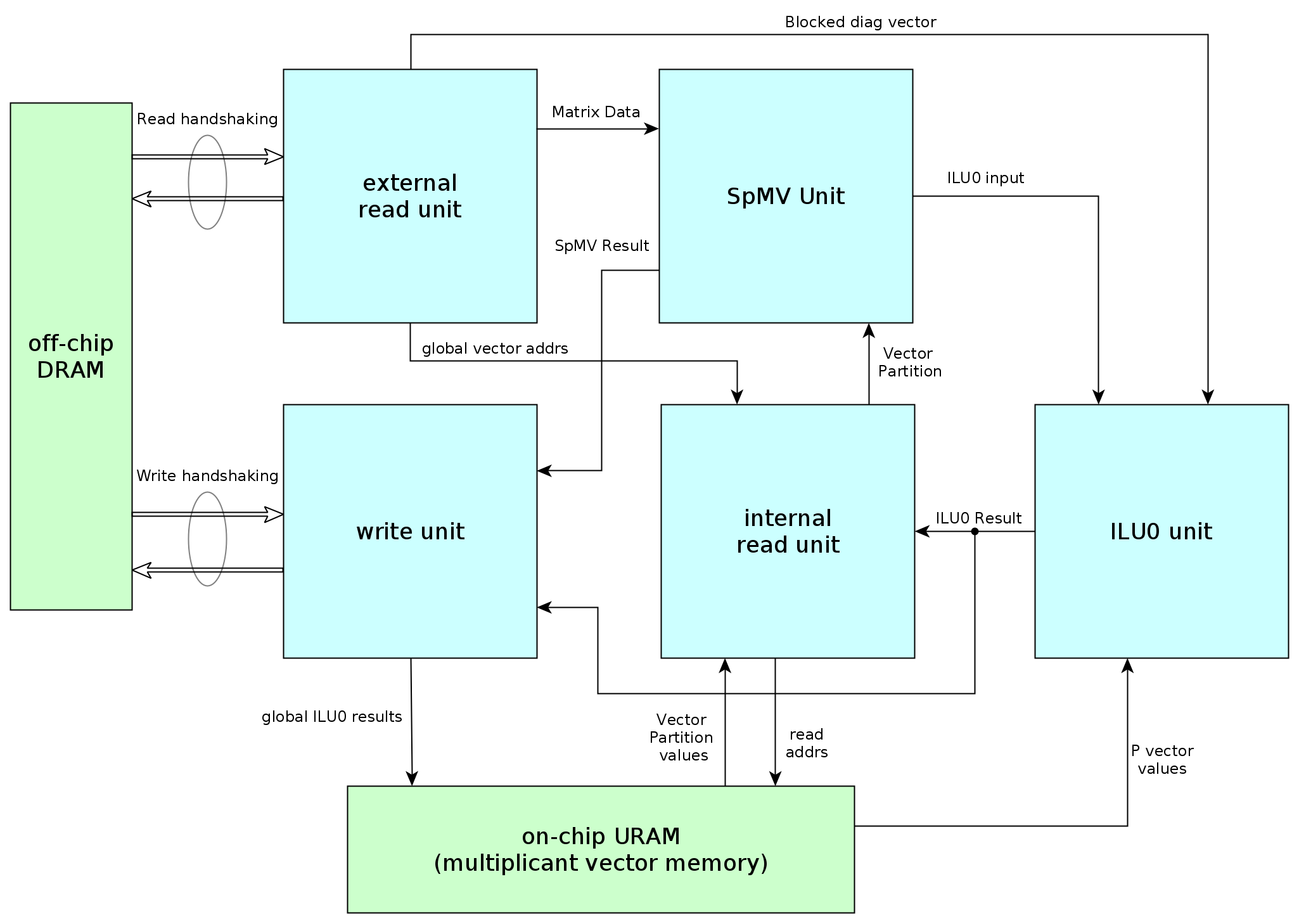}
    \caption{Schematic of the SpMV/ILU0 unit. }
    \label{fig:spmv_ilu0_schematic}
\end{figure}

There are additional operations that the forward and backward substitution needs to apply over the SpMV. For the forward substitution, this is a subtraction of the SpMV unit result from the p vector, and for the backward substitution, it is this subtraction followed by a division by the diagonal value. The ILU0 unit is added to the SpMV unit to perform these operations. The reading of the P vector can be done directly from the URAM, since that is where the P vector is stored during the ILU0 application. The diagonal vector is read from off-chip memory by the external read unit. Figure \ref{fig:spmv_ilu0_schematic} shows how the original SpMV top-level was modified to allow the unit to also apply the ILU0 preconditioning step. Note that this unit is not yet the top-level of the complete solver. 

\subsection{The Vector Operations}

\begin{figure}[!h]
    \centering
    \includegraphics[width=0.4\columnwidth]{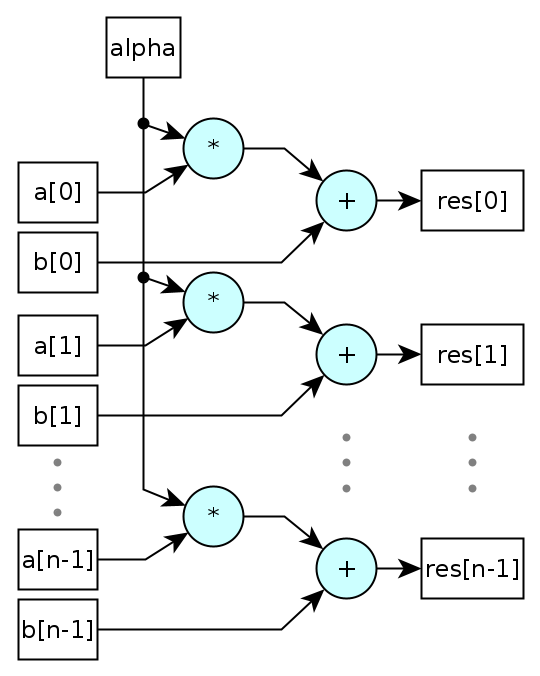}
    \caption{Schematic Overview of dot\_axpy Unit in axpy Mode.}
    \label{fig:axpy_schematic}
\end{figure}

Over the course of a BiCGSTAB run, a number of vector operations of three different kinds need to be applied:
\begin{itemize}
    \item A dot product, which calculates the inner product of two vectors.
    \item An axpy, which performs $\alpha * \vec{x} + \vec{y}$.
    \item A norm, which calculates the absolute value of a vector (i.e. $\sqrt{\vec{x} * \vec{x}}$)
\end{itemize}
Of these, the dot product and norm are very similar in structure, as a norm is the square root of an inner product of a vector with itself. All three operations are highly parallel, and perform the same basic operations (one multiplication and one addition) for every element in their input vectors. For this reason, and to provide more flexibility in exploiting the task-level parallelism in the solver, we designed a unit that can perform either an axpy or a dot product. This dot\_axpy unit contains an equal number of multiply and add floating point units, which can be connected in different ways, depending on which function needs to be performed. When an axpy need to be performed, the adders and multipliers are connected in parallel pipelines that each perform the axpy on one element of the input vectors at a time. For this function, an additional input port is used for the scaling constant $\alpha$, besides the two input vector and one output vector ports. Figure \ref{fig:axpy_schematic} gives an overview of how the adders and multipliers are arranged in the dot\_axpy unit when it performs an axpy operation.

\begin{figure}[!h]
    \centering
    \includegraphics[width=0.6\columnwidth]{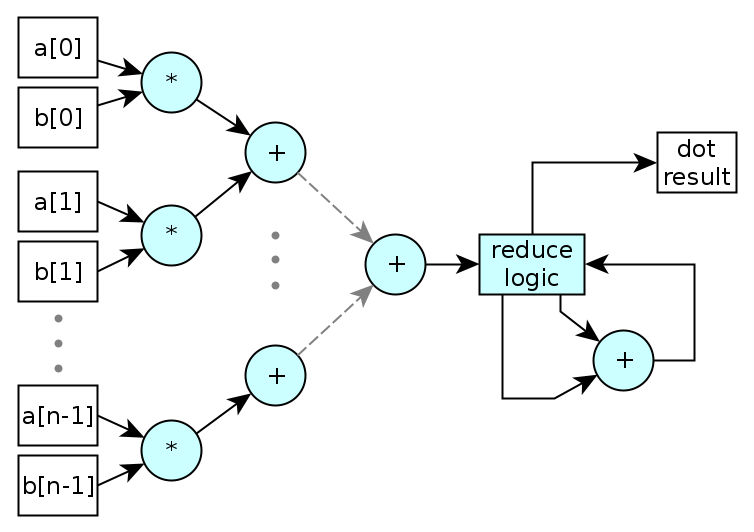}
    \caption{Schematic overview of dot\_axpy unit in dot mode.}
    \label{fig:dot_schematic}
\end{figure}

For the dot products, all multipliers work in parallel, the results of which are then added together into a single value by a tree of adders. The result of this adder tree is added together with the result of previous cycles in the final adder, which adds a new input to its most recent output. After the adder tree has processed all values of the input vectors, logic around the final adder continuously feeds two subsequent valid output of that adder back into it, until one singular finally result is left. Figure \ref{fig:dot_schematic} gives an overview of how the adders and multipliers are arranged in the dot\_axpy unit when it performs a dot product.

During the vector operations, a degree of task-level parallelism exists in the BiCGStab solver algorithm. This parallelism can be found in the following operations (using line numbers from algorithm \ref{alg:BiCGSTAB}): 

\begin{itemize}
    \item The vector subtraction in line 1 can be pipelined with the SpMV operation in that same line, so that the subtraction of each element of $\vec{b}$ happens as soon as the SpMV result in the same position as that element is done. The dot product in line 5 can be pipelined with the subtraction of line 1 in the same way.
    \item The two axpy operations in line 9 can be pipelined.
    \item The dot product in line 12 can be pipelined with the SpMV in line 11.
    \item The axpy operations in lines 14 and 15 can be done in parallel, and the dot product in line 16 can be pipelined with the axpy in line 15.
    \item The two dot products in line 19 can be done in parallel, and can both be pipelined with the SpMV in line 18.
    \item The axpy operations in lines 20 and 21 can be done in parallel. The dot product in line 22 can be pipelined with the axpy in line 21, and so can the dot product in line 7 in the following iteration.
\end{itemize}

Together, these instances of possible task-level parallelism cover all vector operations that the BiCGStab performs. The number of vector operations that can be performed in parallel in each of these instances is, in order: 2, 2, 1, 3, 2 and 4. The complete vector unit contains multiple dot\_axpy units to exploit this task-level parallelism, but to have the dot\_axpy units that it doesn't need to be idle too often, and to not increase the number of read/write ports too much (see section \ref{sec:rw_ports}), we choose to implement 2 dot\_axpy unit in the vector\_ops unit. 

\subsection{Read and Write Ports}
\label{sec:rw_ports}

The Xilinx Alveo U280 data center accelerator card has two different types of large memory that the FPGA chip is connected to: an off-chip DDR memory that consists of two 16 GB DDR4 memory banks, each of which is connected to the FPGA via a single read/write port, and an 8 GB on-chip HBM memory stack that is connected to the FPGA via 32 read/write ports. Each memory port is accessed at the kernel level either via a 512-bit wide AXI4 interface (for the DDR memory) or via a 256-bit wide AXI3 interface (for the HBM memory). This memory architecture has led to the following design decisions:
\begin{itemize}
    \item Reading and writing delays from/to the HBM are slightly shorter than those from/to the DDR memory, so all vectors (which all need to be read and written multiple times during the solver run) are stored in the HBM memory.
    \item To reduce the number of memories to which the host CPU needs to transfer its data, and to thereby reduce the initial data transfer time, all matrix data and the initial B vector are stored on the DDR memory.
    \item Some vector operations read from and write to the same vector, for example the axpy in line 14 of algorithm \ref{alg:BiCGSTAB}. To avoid conflicts and to make data buffering easier, each AXI port is used either to read or write, hence the x, r, and p vectors are stored in two memory locations connected to different ports. During an operation to update one of those vectors, it is read from the port that it was written to most recently, and written to the other port. This essentially makes use of the ping pong buffering technique used to stream data efficiently.
    \item All HBM ports are located on one of the two narrow sides of the FPGA chip (being the shape rectangular), where the HBM crossbar and memory stack is located, so, to avoid routing congestion, we choose to limit the number of HBM ports. We chose to use six HBM ports, because this number of read and write ports can be efficiently used during the vector operations.
    \item Since the width of the read ports is 512 bits, we design our computation pipelines to be able to accept one full cache line every cycle. For the HBM ports, which are natively 256 bit wide, we let the implementation tools to handle the AXI port width conversion. That means each dot\_axpy unit has 8 double-precision multipliers and adders.
\end{itemize}

\subsection{The Complete FPGA Solver}

\begin{figure}[!h]
    \centering
    \includegraphics[width=0.9\columnwidth]{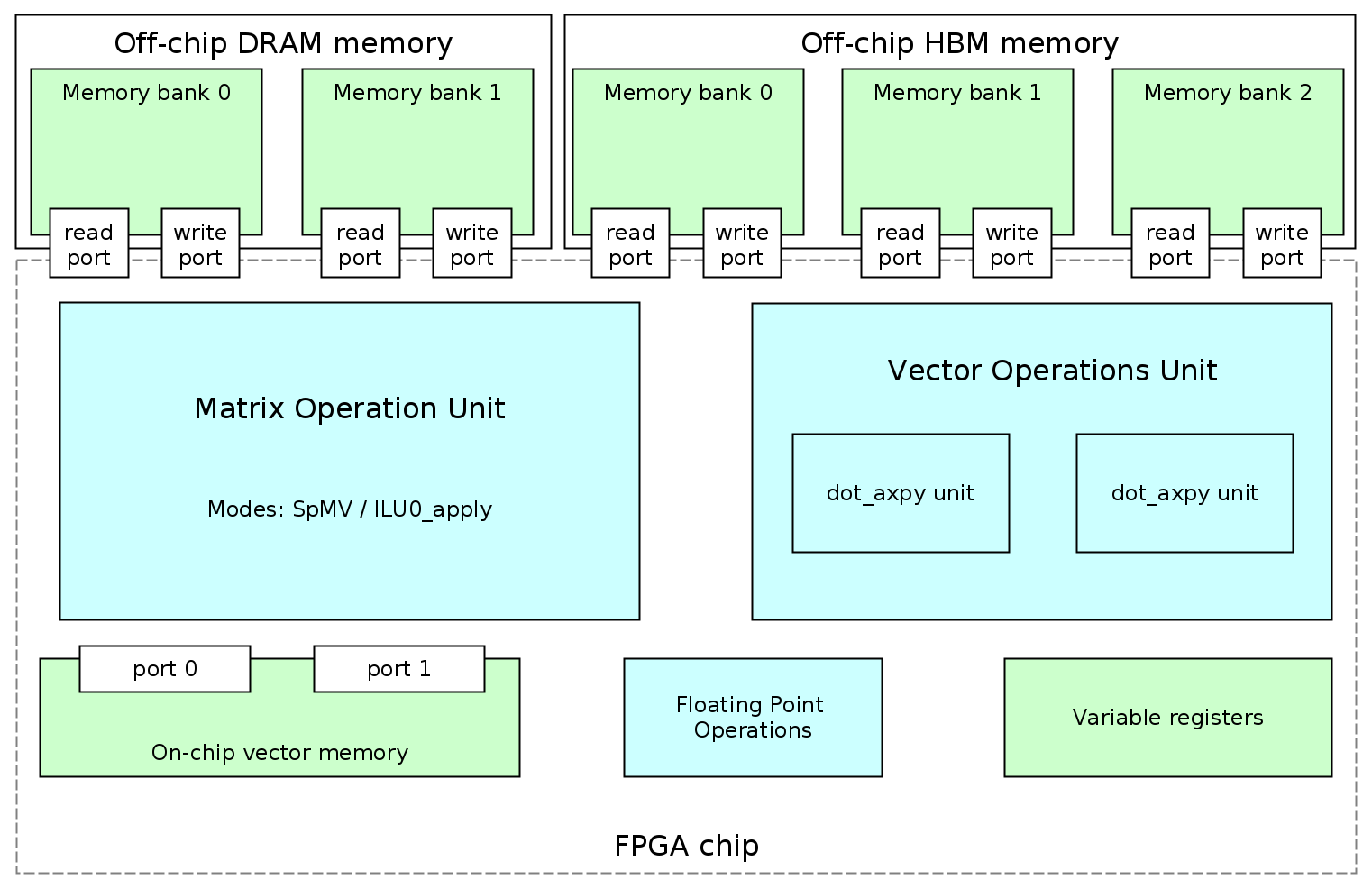}
    \caption{Schematic Overview of the Complete Solver Design.}
    \label{fig:solver_overview}
\end{figure}

A top level solver unit was designed to encompass both the SpMV/ILU0 unit and vector\_ops unit. This top-level unit instructs those units which tasks to perform and with which input/output vectors. It also regulates the access of both units to the URAM internal vector memory and handles filling of this memory during the initialization step of the solver. Figure \ref{fig:solver_overview} gives a schematic overview of the top-level unit of the BiCGStab solver kernel. To not overly complicate the figure, the signals between the units and the memory are not pictured. Besides the units already covered in the previous sections, the figure also shows a block of floating point operators and a memory for floating point variables. The operators are used to apply the operations that only need to be performed on a single value at a time, such as the square root of the norm calculation or the multiplications and division of the $\beta$ calculation in line 8 of algorithm \ref{alg:BiCGSTAB}. The memory is used to stored the results of these operations, and contains $\alpha$, $\beta$, $\omega$, $\rho$, $\rho_{new}$ and conv\_treshold. 

\subsection{The GPU Solvers}
For completeness of the evaluation and comparison of our work, we also developed a couple of GPU solvers using first CUDA and the NVIDIA's cuSPARSE library \cite{cusparse}. cuSPARSE has functions for vector operations like dot, axpy and norm, as well as a blocked ILU0 preconditioner. Second, we also implemented the solver using OpenCL \cite{opencl}, where a similar structure to the CUDA is used, defining kernels for vector operations and applying the ILU0 preconditioner. One big difference is that the construction of the ILU0 preconditioner (the decomposition) is done on the CPU. This has not been implemented on GPU yet, whereas CUDA features an ILU0 decomposition on the GPU in cuSPARSE.

\section{Performance Modeling}
\label{sec:perfmodel}

To predict the performance of both the hardware design as it was in development as well as to gauge how well the design scales, we constructed a performance model that we run in Matlab. This model was made to emulate the data flowing through the computation units and to calculate how many clock cycles the execution would take. This section describes first what parts of the design are modeled and which are not, showing second the predicted modelling results.  

\subsection{Model Details}

The performance model is composed of functions that model the individual hardware kernels: it has a function to estimate the performance of the SpMV, one for the ILU0, and one for the vector operations. To make the performance estimations as accurate as possible, the hardware models work on the same data that is sent to the kernel, including the partitioning data. The exception being that, since the model does not actually perform the solving calculation, it does not read the non-zero matrix values or any vector data. Nevertheless, the partitioning and sparsity pattern information are needed to calculate the performance. Furthermore, the functions that calculate the performance of the ILU0\_apply and SpMV operations share the same structure, in the same way that those operations share the same module in the hardware kernel. These functions do the following for every color:
\begin{itemize}
    \item First, it calculates the time to transfer the vector partition data into the SpMV unit. It does this by dividing the number of values in the partition by the number of ports allocated to the internal URAM memory from which the partition is read.
    \item If apply\_ILU0 is performed, the function calculates the time the transfer of the P vector into the ILU0 unit memory would take in the same manner.
    \item Then, it calculates the time the SpMV pipeline takes. During this step, the function's emphasis is on the time the reading of the matrix data takes. It assumes a set bandwidth for each of three ports from which the three matrix arrays (non-zero values, column indices and new row offsets) are read, and simulates reading into three FIFOs at the speed set by that bandwidth. Whenever enough data is available in the FIFOs for all three arrays to fill an input line of the SpMV pipeline, that data is removed from the FIFOs.
    \item After enough data has been read from all three ports for the current color, and the simulated FIFOs are empty, a set delay is added for the SpMV pipeline to write back the result data. Since much less data needs to be written back than needed to be read, no delay beyond the regular write overhead of the final lines is deemed necessary. 
    \item If apply\_ILU0 is performed, a set delay for the ILU0 unit is added, and the write delay is calculated with the on-chip URAM instead of an off-chip memory.
\end{itemize}

The cycle count results of all colors are added together to find the total number of clock cycles that the SpMV or apply\_ILU0 takes. The calculations done to estimate the cycle count of the vector operations are comparatively more simple, but follow the same method: first the reading of data into FIFOs at a set bandwidth is modeled, data is removed from the FIFO whenever enough is in it to fill up a line of inputs of the vector operation unit, and finally, when all reading is done, the latency of the pipeline of the operation is added, as well as some writing overhead cycles for the axpy operation.\\
A top-level function adds together the cycle count results of apply\_ILU0, SpMV, and vector operations to get to a final estimate. This top-level function has a number of variables that can be changed to do domain space exploration and to assess the scalability of the kernel. These are: 
\begin{itemize}
    \item The number of multipliers and adders in both the vector operation and SpMV/ILU0 units.
    \item The bandwidth to the off-chip memory in GB/s.
    \item The bandwidth to the on-chip URAM in number of ports (how many values of the vector stored on-chip can be read simultaneously).
    \item The delay of the adder and multiplier units
    \item The matrix that is modeled for, and how long the solve of that matrix takes, in iterations.
\end{itemize}
Please note that the largest source of performance uncertainty in the design is the time the memory operations take. Memory accesses don't happen at a constant pace, but in bursts, the size of which and the time between them may very based on various factors related to the memory. As a result, the weakest part of the model is that it does not accurately capture this memory behaviour. Similarly, it does not support modeling reading two arrays of data from the same port, nor the performance penalties from reading and writing on the same memory bank at the same time. 

\begin{figure*}
    \centering
    \includegraphics[width=\textwidth]{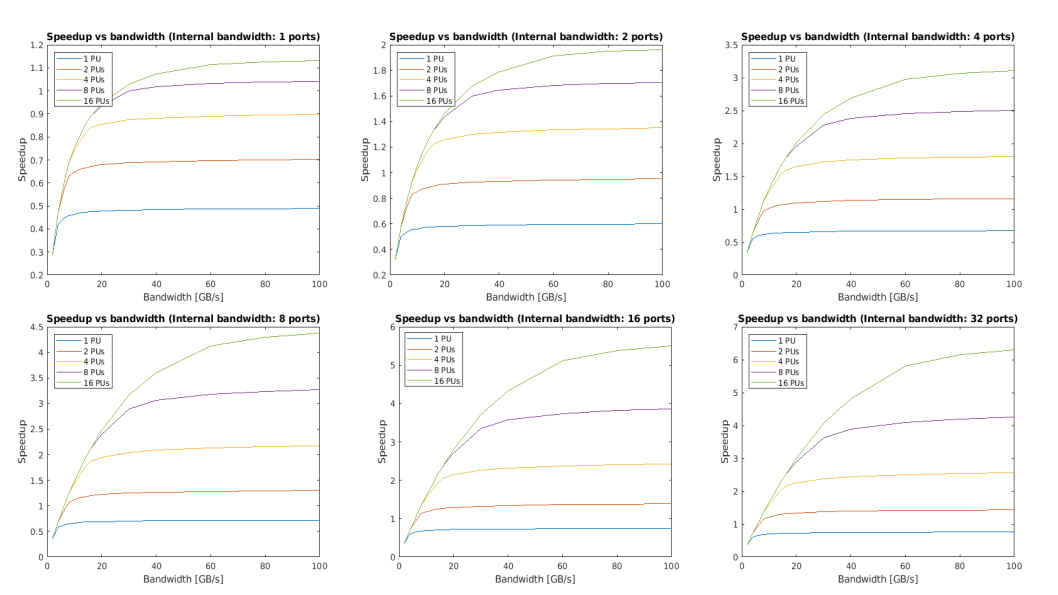}
    \caption{Domain Search Exploration for the ILU0-BiCGStab Solver. One internal port has a maximum bandwidth of around 18 GB/s at the current kernel frequency of 280 MHz. The configuration achieved in this paper for the FPGA solver was 2 internal ports, 8 PUs, and 50 GB/s external bandwidth. This corresponds to the point in the top middle figure on the parse line estimating a 1.6x speedup, which was confirmed by the experimental results presented next. %\notern{TODO-1:  draw a point in the sub-figure and the proper PU line where we comment within a textbox that that is the configuration of the achieved design in this paper! Subsequently, connect to the story that there is room for scaling. TODO-2: Figure is old and compares against an old CPU baseline (also single-thread?). Comment on that! Where is our design when comparing to the newer CPU, i.e., we should compare against 4 threads nimbix CPU! TODO-3: maybe regenerate the figures in octave to reflect the newer CPU and our latest hw design?}}
    }
    \label{fig:DSEsolver}
\end{figure*}

\subsection{Modelling Results}

%\notern{
We use the model to analyze the scalability of the design and to understand what are the critical points in achieving the best performance. We perform a domain space exploration by varying the model parameters, starting with the number of adders and multipliers set at 2, increasing the bandwidth from 10 GB/s, and increasing the number of internal ports from 2. When a parameter does not give anymore a performance increase, we switch to increase the new bottleneck parameter. We perform this exercise by increasing all parameters until a maximum value we select as an upper bound of a maximum practical boundary for our design. Please note that even the maximum theoretical bandwidth is 460 GB/s, due to the complexity of the design which influence the mapping and routing of resources from/to the memory, we restrict the bandwidth to only 100 GB/s as parameter. Figure \ref{fig:DSEsolver} shows the modelling results.
%}. 

%\notern{TODO-11: Figure \ref{fig:DSEsolver} shows the modelling results could be added here. Which ones?: Add the graphs you included at some point in a presentation, in which you were showing even a 6x speedup by varying the parameters. I believe there were 6 figures in 1 slide. Conclude that the design shows good scaling potential, which thus validates the design and therefore we implemented it as described in the previous section. We now move to analyze how much of this scaling can be achieved in practice, which we present next.}

\begin{table*}[!h]
\caption{Characteristics of the platforms used to benchmark the kernels}
\begin{center}
\label{tab:platforms}
\begin{tabular}{|l|c|c|c|c|c|c|}
\hline
\multirow{2}{*}{\begin{tabular}[c]{@{}c@{}}Short\\ name\end{tabular}} &
     \multicolumn{3}{c|}{Host} & \multicolumn{2}{c|}{Accelerator} \\
\cline{2-6}
 & CPU (\#Total cores) & Frequency & Memory (\#Channels / Bandwidth) & Name & Memory (Bandwidth) \\
\hline
\emph{FPGA} & Xeon E5-2640V3 (16) & 2.6 GHz & DDR4-2133 (4 / 59 GB/s) & Xilinx Alveo U280 & DDR4+HBM (498 GB/s) \\
\emph{GPU$^1$} & Xeon E5-2640V3 (16) & 2.6 GHz & DDR4-2133 (4 / 59 GB/s) & Nvidia K40m & GDDR5 (288 GB/s)\\
\emph{GPU$^2$} & Xeon E5-2698V4 (40) & 2.2 GHz & DDR4-2400 (4 / 77 GB/s) & Nvidia V100-SXM2-16GB & HBM2 (900 GB/s)\\
\emph{CPU}  & Xeon E5-2698V4 (40) & 2.2 GHz & DDR4-2400 (4 / 77 GB/s) & -- & -- \\
\hline
\end{tabular}
\end{center}
\end{table*}

\section{Experimental Results}
\label{sec:results}

Evaluation of the performance of our design is done in three stages. Firstly, the performance of the stand-alone SpMV unit is evaluated, followed by the performance of the complete solver stand-alone. Finally, the FPGA solver is integrated into the flow simulator, and its performance is compared to that of a CPU and a GPU. All the FPGA bitstream implementations were done in Xilinx Vitis 2019.2 and run on a Xilinx Alveo U280 data center accelerator card \cite{u280}. In this section, performance for the SpMV unit will be reported in Giga FLoating-point OPerations per second. These are calculated by dividing the total number of FP operations performed by the time it took to perform them. Therefore, the total number of required operations is two times the number of non-zero values in the matrix that is being multiplied (since one multiplication and one addition need to be performed on each non-zero value).\\
Table \ref{tab:platforms} lists the platforms we used to benchmark the kernels and the flow simulator. Each platform will be then referred in the rest of the paper with its short name (e.g. \emph{FPGA}).

\subsection{SpMV results}

\begin{figure*}[!h]
    \centering
    \includegraphics[width=1.0\textwidth]{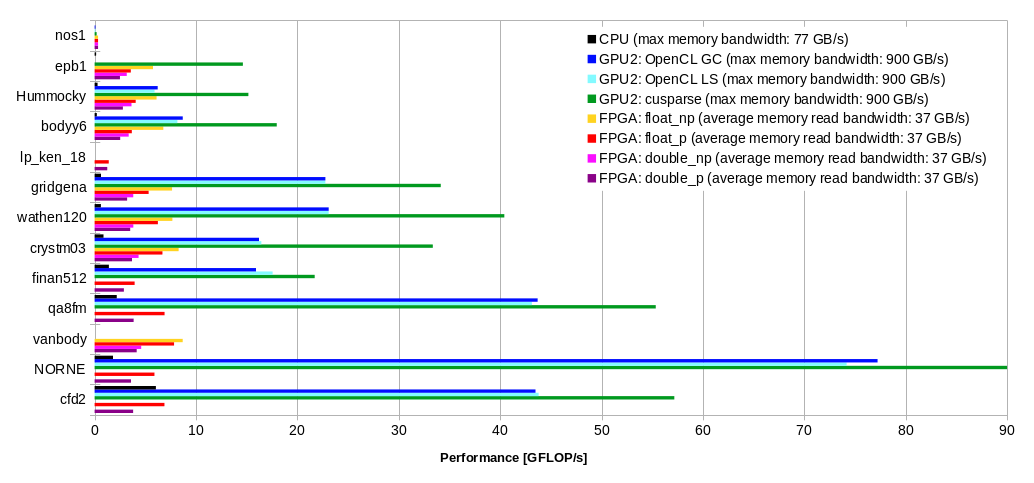}
    \caption{SpMV performance. Please note that for the FPGA, the maximum aggregated HBM memory bandwidth is 460 GB/s, however, due to design limitations caused by implementation issues, the maximum bandwidth used by the kernel is around 50 GB/s. }
    \label{fig:spmv_GFLOPs}
\end{figure*}

The SpMV unit has many configurable parameters that will impact both performance and FPGA resource usage. For this performance evaluation, we choose to compare between single-precision (float) and double-precision (double) implementations, as well as between partitioned (\_p) and non-partitioned (\_np) implementations. Other parameters were set to the values in Table \ref{tab:spmv_params}.

\begin{table}[!h]
\caption{SpMV parameters of different SpMV implementations}
\begin{center}
\label{tab:spmv_params}
\begin{tabular}{|l|c|c|c|c|}
\hline
Parameter & double\_p & double\_np & float\_p & float\_np  \\ 
\hline
Num mults & 8 & 8 & 16 & 16 \\
Max rows & 262144 & 65536 & 262144 & 65536 \\
Read ports & 4 & 4 & 4 & 4 \\
\hline
\end{tabular}
\end{center}
\end{table}

The resource utilization of the four chosen SpMV kernel implementations is shown in Table \ref{tab:spmv_utilization}. From this table, we can observe that the BRAM utilization sets the limit of how often this kernel could be replicated on this same FPGA device. All implementations are able to reach a frequency of 300 MHz, which is the maximum frequency at which the AXI memory interfaces can run.

\begin{table}[!h]
\caption{FPGA resource utilization of different implementations of the SpMV kernel as \% of the resources available in the FPGA's dynamic region}
\begin{center}
\label{tab:spmv_utilization}
\begin{tabular}{|l|r|r|r|r|}
\hline
Resource & double\_p & double\_np & float\_p & float\_np  \\ 
\hline
LUT & 4.08 \% & 4.12 \% & 4.92 \% & 5.00 \% \\
LUTRAM & 0.70 \% & 0.69 \% & 0.82 \% & 0.78 \% \\
REG & 4.14 \% & 4.45 \% & 4.96 \% & 5.24 \% \\
BRAM & 15.27 \% & 12.38 \% & 18.97 \% & 12.38 \% \\
URAM & 4.17 \% & 6.67 \% & 2.50 \% & 6.67 \% \\
DSP & 0.93 \% & 0.93 \% & 0.74 \% & 0.74 \% \\
\hline
\end{tabular}
\end{center}
\end{table}

The sizes and types of matrices that were used to test the SpMV unit are listed in Table \ref{tab:spmv_matrices}. All of the listed matrices were obtained from the SuiteSparse matrix collection \cite{suitesparse, suitesparsepaper}, apart for the NORNE matrix, which models a real-life oil field, and is a testcase in the OPM framework. As such, this matrix was not obtained from the SuiteSparse matrix collection, but instead by exporting it from the flow simulator. Please note that all but one of the matrices are square, and of the square matrices, three of them are non-symmetric. These matrices were chosen to prove that the SpMV unit works on such non-symmetric and non-square matrices. Some of these matrices are not used as benchmarks for the BiCGSTAB solver, however, as it does not work for non-square matrices and has additional limitations as explained in the next section. 

\begin{table} [!h]
\caption{Characteristics of the matrices used to benchmark the SpMV kernel}
\begin{center}
\label{tab:spmv_matrices}
\begin{tabular}{|c|c|c|c|c|c|}
\hline
Name & Rows & Columns & NNZs & Density\% & Symm. \\ 
\hline
nos1 & 237 & 237 & 2115 & 1.811 & yes  \\ 
epb1 & 14734 & 14734 & 95053 & 0.044 & no \\
Hummocky & 12380 & 12380 & 121340 & 0.079 & yes \\
bodyy6 & 19366 & 19366 & 134748 & 0.036 & yes \\
lp\_ken\_18 & 105127 & 154699 & 358171 & 0.002 & no \\
gridgena & 48962 & 48962 & 512084 & 0.021 & yes \\
wathen120 & 36441 & 36441 & 565761 & 0.043 & yes \\
finan512 & 74752 & 74752 & 596992 & 0.011 & yes \\
qa8fm & 66127 & 66127 & 1660579 & 0.038 & yes \\
crystm03 & 24696 & 24696 & 1751310 & 0.096 & yes \\
vanbody & 47072 & 47072 & 2336898 & 0.105 & yes \\
NORNE & 133293 & 133293 & 2818879 & 0.016 & no \\
cfd2 & 123440 & 123440 & 3087898 & 0.020 & yes \\
\hline
\end{tabular}
\end{center}
\end{table}

In Figure \ref{fig:spmv_GFLOPs}, the performance for all the matrices used to benchmark the kernel are shown. The corresponding tabulated values can be found in Table \ref{tab:spmv_performance}. For the FPGA and GPU, those performance numbers do not include the data transfer time between the host CPU and the on-board memories. From this figure, we can see that the single-precision versions of the FPGA kernel achieve almost 2 times more FLOP/s than their double-precision counterparts. This can be explained by the fact that the amount of data that needs to be read for each single-precision operations is half of the amount of data that needs to be read for a double-precision one. We also observe, as expected, that the non-partitioned version achieves a slightly higher performance than the corresponding partitioned version, at the trade-off that the non-partitioned version cannot operate on the largest of the benchmark matrices.

\begin{table*}[!h]
\caption{Performance in GFLOP/s for the SpMV kernel}
\begin{center}
\label{tab:spmv_performance}
\begin{tabular}{|l|c|c|c|c|c|c|c|c|}
\hline
\multirow{2}{*}{Matrix} & \multicolumn{4}{c|}{FPGA} & \multicolumn{3}{c|}{GPU$^2$} & CPU \\
\cline{2-8}
 & double\_p & double\_np & float\_p & float\_np & cusparse & OpenCL LS & OpenCL GC & (double) \\
\hline
nos1 & 0.34 & 0.34 & 0.34 & 0.34 & 0.20 & 0.08 & 0.08 & 0.00 \\
epb1 & 2.50 & 3.17 & 3.57 & 5.76 & 14.62 & -- & -- & 0.15 \\
Hummocky & 2.79 & 3.64 & 4.05 & 6.11 & 15.17 & 5.92 & 6.22 & 0.29 \\
bodyy6 & 2.52 & 3.36 & 3.67 & 6.78 & 17.97 & 8.17 & 8.69 & 0.22 \\
lp\_ken\_18 & 1.25 & -- & 1.40 & -- & -- & -- & -- & -- \\
gridgena & 3.21 & 3.81 & 5.33 & 7.64 & 34.14 & 22.76 & 22.76 & 0.63 \\
wathen120 & 3.51 & 3.82 & 6.24 & 7.67 & 40.41 & 23.09 & 23.09 & 0.62 \\
crystm03 & 3.69 & 4.34 & 6.69 & 8.29 & 33.36 & 16.44 & 16.22 & 0.88 \\
finan512 & 2.89 & -- & 3.95 & -- & 21.71 & 17.56 & 15.92 & 1.41 \\
qa8fm & 3.85 & -- & 6.90 & -- & 55.35 & 43.13 & 43.70 & 2.18 \\
vanbody & 4.15 & 4.60 & 7.84 & 8.69 & -- & -- & -- & -- \\
NORNE & 3.59 & -- & 5.90 & -- & 90.93 & 74.18 & 77.23 & 1.81 \\
cfd2 & 3.80 & -- & 6.89 & -- & 57.18 & 43.80 & 43.49 & 6.04 \\
\hline
\end{tabular}
\end{center}
\end{table*}

\subsection{Stand-alone FPGA solver results}

For the ILU0 BiCGSTAB solver, we are only interested in a version of the solver that can run on larger matrices with high precision, so only the double-precision, partitioned version of the solver is benchmarked. The design parameters as listed for the double\_p SpMV implementation in Table \ref{tab:spmv_params} hold for this solver implementation as well, except for the number of read ports, which is increased by one. \\
In Table \ref{tab:solver_utilization}, the resource utilization of the ILU0 BICGSTAB solver is compared to that of the SpMV kernel with the same parameters. In both cases, these values refer to the percentage of resources used over the available ones in the dynamic region (which is reconfigurable by the user), and they include all the supporting circuitry to control the solver and access the memory on the FPGA board. Hence, not included in these values are the resources used by the FPGA's static region, which is provided by Xilinx. Due to the increased complexity of the solver over the SpMV kernel, it could only reach a frequency of 280 MHz.

\begin{table}[!h]
\caption{FPGA resource utilization of the solver compared to the SpMV kernel as \% of the resources available in the FPGA's dynamic region}
\begin{center}
\label{tab:solver_utilization}
\begin{tabular}{|l|r|r|}
\hline
Resource & SpMV & Solver \\ 
\hline
LUT & 4.08 \% & 6.93 \% \\
LUTRAM & 0.70 \% & 1.13 \% \\
REG & 4.14 \% & 6.41 \% \\
BRAM & 15.27 \% & 24.64 \% \\
URAM & 4.17 \% & 4.17 \% \\
DSP & 0.93 \% & 3.18 \%  \\
\hline
\end{tabular}
\end{center}
\end{table}

Some of the matrices used to test the SpMV unit were not usable during the testing of the solver, because they were non-square, had zero values on their diagonals, or exceeded on-board memories of the design. Table \ref{tab:solver_matrices} lists the matrices that were used to test the BICGSTAB solver, along with their original sizes and the size of the derived L/U matrices used by the FPGA solver.

\begin{table}[!h]
\caption{Characteristics of the testing matrices}
\begin{center}
\label{tab:solver_matrices}
\begin{tabular}{|c|c|c|c|c|c|}
\hline
\multirow{3}{*}{Name} & \multirow{3}{*}{Dim.} & \multicolumn{4}{c|}{Non-zeroes} \\
\cline{3-6}
 & & \multirow{2}{*}{\begin{tabular}[c]{@{}c@{}}FPGA\\ matrix\end{tabular}} &
     \multirow{2}{*}{\begin{tabular}[c]{@{}c@{}}FPGA\\ L-matrix\end{tabular}} &
     \multirow{2}{*}{\begin{tabular}[c]{@{}c@{}}FPGA\\ U-matrix\end{tabular}} &
     \multirow{2}{*}{\begin{tabular}[c]{@{}c@{}}GPU\\ matrix\end{tabular}} \\
 & & & & & \\
\hline
nos1 & 237 & 1017 & 390 & 390 & 1017 \\ 
bodyy6 & 19366 & 134197 & 121414 & 70762 & 134197 \\
gridgena & 48962 & 513060 & 421068 & 315336 & 513060 \\
crystm03 & 24696 & 583770 & 279519 & 279522 & 583770 \\ 
wathen120 & 36441 & 565761 & 468529 & 379446 & 565761 \\
qa8fm & 66127 & 1660564 & 1431908 & 1251238 & 1660564 \\
NORNE & 133293 & 1314999 & 726369 & 513512 & 2818879 \\
\hline
\end{tabular}
\end{center}
\end{table}

\begin{figure*}[!h]
    \centering
    \includegraphics[width=1.0\textwidth]{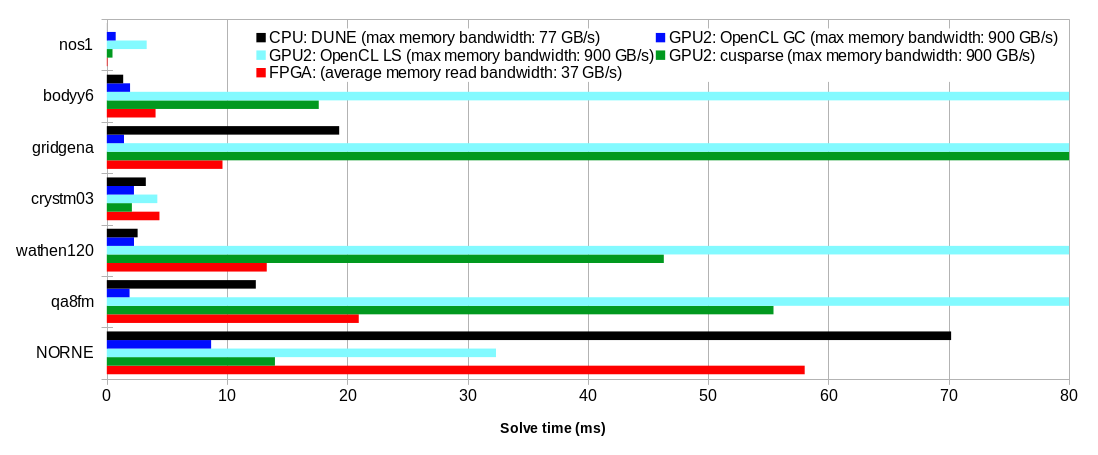}
    \caption{Solver execution time for selected matrices. For readability of the CPU/FPGA results, some GPU results have been clipped. Refer to Table \ref{tab:solver_run_times} for the actual values. Please note that for the FPGA, the maximum aggregated HBM memory bandwidth is 460 GB/s and the maximum aggregated DDR4 memory bandwidth is around 37 GB/s, however, due to design limitations caused by implementation issues, the maximum bandwidth used by the kernel is around 52 GB/s.}
    \label{fig:solver_performance}
\end{figure*}

The performance results obtained after running the stand-alone ILU0 BiCGSTAB solver are shown in Figure \ref{fig:solver_performance}. For each matrix, we present the best result obtained either by using the graph-coloring or the level-scheduling reordering algorithms. The detailed timings are listed in Table \ref{tab:solver_run_times}, which contains the run time as reported by the profiling info (\emph{Solver}) of the runs (only for the FPGA, used in Figure \ref{fig:solver_performance}), the run time as measured by the \emph{Host} (used in the figure for the CPU and GPU results), and the time spent transferring (\emph{Transfer}) the input data and the output results between the host CPU's and the accelerators' on-board DRAM/HBM memories. Also, the number of iterations (\emph{Iter\#}) needed to solve the system are reported. The \emph{Host} time is, in all cases, higher than the \emph{Solver} time, as the former includes overhead of starting and waiting for the end of the solver kernel. The transfer times are not included in the other two times.

\begin{table*}[!h]
\caption{Timing results (in milliseconds) for solving selected matrices}
\begin{center}
\label{tab:solver_run_times}
\begin{tabular}{|l|c|c|c|c|c|c|c|c|c|c|c|c|c|c|c|}
\hline
\multirow{3}{*}{Matrix} & \multicolumn{4}{c|}{ \multirow{2}{*}{FPGA} } & \multicolumn{2}{c|}{ \multirow{2}{*}{CPU DUNE} } & \multicolumn{9}{c|}{GPU$^2$} \\
\cline{8-16}
& \multicolumn{4}{c|}{} & \multicolumn{2}{c|}{} & \multicolumn{3}{c|}{cusparse} & \multicolumn{3}{c|}{OpenCL LS} & \multicolumn{3}{c|}{OpenCL GC} \\
\cline{2-16}
& Solver & Host & Transfer & Iter\# &
  Host & Iter\# &
  Host & Transfer & Iter\# &
  Host & Transfer & Iter\# &
  Host & Transfer & Iter\#\\
\hline
nos1      & 0.1 & 0.4 & 0.4 & 1.0    & 0.0 & 0.5  & 0.5 & 0.0 & 0.5   & 3.3 & 0.1 & 1.0    & 0.7 & 0.1 & 1.5 \\
bodyy6    & 4.1 & 4.3 & 3.2 & 2.0    & 1.4 & 0.5  & 17.6 & 0.5 & 0.5  & 136.5 & 0.5 & 1.0  & 1.9 & 0.5 & 2.0 \\
gridgena  & 9.6 & 10.2 & 12.4 & 1.5  & 19.3 & 4.5 & 528.7 & 1.2 & 4.5 & 1007.7 & 1.3 & 2.5 & 1.4 & 1.3 & 1.5 \\
crystm03  & 4.4 & 4.9 & 9.4 & 0.5    & 3.2 & 0.5  & 2.1 & 1.6 & 0.5   & 4.2 & 1.7 & 0.5    & 2.3 & 1.7 & 1.0 \\
wathen120 & 13.3 & 13.5 & 12.1 & 2.0 & 2.6 & 0.5  & 46.3 & 1.1 & 0.5  & 334.3 & 1.3 & 1.0  & 2.3 & 1.3 & 2.0 \\
qa8fm     & 20.9 & 21.2 & 30.6 & 1.0 & 12.4 & 0.5 & 55.4 & 3.4 & 0.5  & 179.6 & 3.6 & 0.5  & 1.9 & 3.7 & 1.0 \\
NORNE     & 58.0 & 58.6 & 27.3 & 4.5 & 70.2 & 4.5 & 14.0 & 2.7 & 4.5  & 32.4 & 3.2 & 4.5   & 8.7 & 3.0 & 6.5 \\
\hline
\end{tabular}
\end{center}
\end{table*}

In Table \ref{tab:solver_bandwidth_util} we report the bandwidth utilization for each memory port as recorded by the profiling facilities present in the FPGA solver. Each column shows a memory port, along with its mapping (either to DDR4 or HBM memory). The bandwidth utilization is given a percentage of the maximum bandwidth available for each individual port, which is 18.8 GB/s for ports read 0-1 (DDR4 memory, 512-bit AXI bus @ 300 MHz), while it's 14.1 GB/s for the remaining ports (HBM memory, 256-bit AXI bus @ 450 MHz).

\begin{table*}
\caption{Read and write ports bandwidth utilization of the FPGA solver, as \% of the available bandwidth}
\begin{center}
\label{tab:solver_bandwidth_util}
\begin{tabular}{|l|c|c|c|c|c|c|c|c|c|c|}
\hline
\multirow{2}{*}{Matrix} & \multicolumn{2}{c|}{DDR4} & \multicolumn{6}{c|}{HBM} & 
  \multirow{2}{*}{\begin{tabular}[c]{@{}c@{}}Aggregated read\\bandwidth (GB/s)\end{tabular}} &
  \multirow{2}{*}{\begin{tabular}[c]{@{}c@{}}\% of max aggregated\\read bandwidth\end{tabular}} \\
\cline{2-9}
& Read 0 & Read 1 & Read 2 & Read 3 & Read 4 & Write 0 & Write 1 & Write 2 &  & \\
\hline
nos1 & 42.8 & 14.9 & 50.6 & 50.5 & 49.5 & 79.4 & 79.4 & 79.1 & 32.0 & 40.2\\ 
bodyy6 & 82.6 & 36.8 & 70.8 & 64.4 & 72.4 & 100 & 100 & 100 & 51.6 & 64.7 \\
gridgena & 81.7 & 36.2 & 70.0 & 65.8 & 70.2 & 100 & 100 & 100 & 51.1 & 64.1 \\
crystm03 & 79.8 & 43.2 & 68.3 & 69.2 & 69.3 & 100 & 100 & 100 & 52.2 & 65.4\\ 
wathen120 & 80.6 & 35.6 & 68.9 & 67.0 & 70.0 & 100 & 100 & 100 & 50.7 & 63.7 \\
qa8fm & 81.4 & 36.0 & 68.6 & 66.5 & 72.1 & 100 & 100 & 100 & 51.1 & 64.2 \\
NORNE & 83.0 & 35.8 & 66.8 & 62.5 & 69.6 & 100 & 100 & 100 & 50.3 & 63.1 \\
\hline
\end{tabular}
\end{center}
\end{table*}

\subsection{Flow results}
\label{sec:flow_results}

The ILU0 BiCGSTAB solvers were integrated into the flow simulator in OPM \cite{opmrepo}, by replacing the call to the simulator's own iterative solver with calls to a function that does the required pre-processing, sends the resulting data to the accelerator, and then reads the results from it after it is done.
The source code used was from version 2020.10-rc4, with custom modifications to integrate our solvers.
The original flow solver is an ILU0 BICGSTAB solver based on the DUNE project \cite{dune}, and in Table \ref{tab:flow_results_all}, the performance results between this software solver and our solvers are compared when running simulation on the NORNE testcase (file \emph{NORNE\_ATW2013.DATA}) \cite{NORNE} that is part of the OPM project.
For all runs, the parameter '--matrix-add-wellcontributions=true' was added to the command line (to have a fair comparison with the FPGA results, because the FPGA solver currently does not support separated well contributions). For the Nornes case, including well contributions will increase run time on the CPU with approximately twenty percent compared to the default configuration. Moreover, for the FPGA run, the parameter '--threads-per-process=8' was also added. This will only impact the assembly part of the simulation, not the preconditioner nor the linear solver, in effect reducing the computational time spent outside the preconditioning and linear solve part. The reordering algorithm has been left to the default for FPGA (level scheduling), because graph coloring produces in this case more colors than the ones supported by the FPGA solver. For the executions using the FPGA solver, all the application's threads were pinned to CPU \#0, i.e. the one directly connected to the FPGA, to avoid fluctuations in bandwidth during the transfers between the host and the FPGA memory.\\
The default reduction of $1e-2$ was used for the results shown in Table \ref{tab:flow_results_all} and Figure \ref{fig:solver_performance}, while a reduction of $1e-6$ was used for the results shown in Table \ref{tab:flow_results_red1e-6_all}. For an explanation of the characteristics of each platform used to run flow, please refer to Table \ref{tab:platforms}.

\begin{table*}
\caption{Comparison of performance for flow with reduction=1e-2(default) using CPU, FPGA and GPU (in seconds)}
\begin{center}
\label{tab:flow_results_all}
\begin{tabular}{|l|c|c|c|c|c|c|c|c|}
\hline
\multirow{2}{*}{Function} & \multirow{2}{*}{CPU} & \multirow{2}{*}{FPGA} & 
 \multicolumn{3}{c|}{GPU$^{1}$} & \multicolumn{3}{c|}{GPU$^{2}$} \\
\cline{4-9}
&&& cusparse & OpenCL LS & OpenCL GC & cusparse & OpenCL LS & OpenCL GC \\
\hline
Total time & 540.6 & 751.8 & 489.7 & 799.6 & 749.3 & 320.1 & 507.6 & 458.5 \\
\hspace{8pt}Solver time & 540.6 & 751.7 & 489.6 & 799.5 & 749.2 & 320.0 & 507.6 & 458.5 \\
\hspace{8pt}Assembly time & 99.4 & 94.6 & 133.7 & 137.5 & 163.6 & 115.6 & 112.7 & 114.7 \\
\hspace{16pt}Well assembly time & 43.0 & 47.1 & 21.15 & 21.6 & 26.5 & 13.9 & 14.6 & 16.8 \\
\hspace{8pt}Linear solve time & 339.5 & 559.9 & 248.9 & 551.4 & 460.6 & 117.6 & 293.6 & 207.8 \\
\hspace{16pt}Linear setup time & 41.8 & 45.3 & 38.1 & 39.5 & 47.7 & 33.6 & 35.4 & 43.6 \\
\hspace{8pt}Update time & 57.3 & 54.4 & 67.3 & 70.1 & 84.5 & 53.6 & 56.9 & 71.2 \\
\hspace{8pt}Output write time & 3.5 & 3.5 & 3.2 & 3.2 & 3.2 & 2.9 & 2.9 & 3.0 \\
\hline
Number of linear solves & 1458 & 1441 & 1437 & 1507 & 1818 & 1449 & 1507 & 1861 \\
Number of linear iterations & 21991 & 21012 & 21282 & 22626 & 42160 & 21020 & 22626 & 40991 \\
\hhline{|=|=|=|=|=|=|=|=|=|}
Speedup total time & 1.00 & 0.72 & 1.10 & 0.68 & 0.72 & 1.69 & 1.07 & 1.18 \\
Speedup linear solver time & 1.00 & 0.61 & 1.36 & 0.62 & 0.74 & 2.89 & 1.16 & 1.63 \\
\hline
\end{tabular}
\end{center}
\end{table*}

\begin{table*}
\caption{Comparison of performance for flow with reduction=1e-6 using CPU, FPGA and GPU (in seconds)}
\begin{center}
\label{tab:flow_results_red1e-6_all}
\begin{tabular}{|l|c|c|c|c|c|c|c|c|}
\hline
\multirow{2}{*}{Function} & \multirow{2}{*}{CPU} & \multirow{2}{*}{FPGA} & 
 \multicolumn{3}{c|}{GPU$^{1}$} & \multicolumn{3}{c|}{GPU$^{2}$} \\
\cline{4-9}
&&& cusparse & OpenCL LS & OpenCL GC & cusparse & OpenCL LS & OpenCL GC \\
\hline
Total time & 1254.1 & 1512.9 & 900.0 & 1893.0 & 1333.3 & 484.5 & 950.5 & 579.4 \\
\hspace{8pt}Solver time & 1254.0 & 1512.9 & 900.0 & 1893.0 & 1333.2 & 484.5 & 950.5 & 579.3 \\
\hspace{8pt}Assembly time & 82.2 & 84.9 & 112.0 & 114.7 & 115.9 & 101.8 & 101.7 & 101.7 \\
\hspace{16pt}Well assembly time & 36.1 & 42.3 & 17.1 & 17.5 & 17.6 & 11.9 & 12.1 & 12.5 \\
\hspace{8pt}Linear solve time & 1081.3 & 1333.8 & 691.4 & 1680.1 & 1119.5 & 303.8 & 768.8 & 397.0 \\
\hspace{16pt}Linear setup time & 35.2 & 38.9 & 31.1 & 31.4 & 31.3 & 27.8 & 28.2 & 28.7 \\
\hspace{8pt}Update time & 46.1 & 50.9 & 56.9 & 57.4 & 57.3 & 45.6 & 45.6 & 45.8 \\
\hspace{8pt}Output write time & 3.5 & 3.5 & 3.0 & 3.2 & 3.2 & 2.8 & 2.9 & 2.9 \\
\hline
Number of linear solves & 1207 & 1207 & 1202 & 1207 & 1207 & 1207 & 1207 & 1207 \\
Number of linear iterations & 79163 & 78275 & 78447 & 82673 & 117463 & 78958 & 82673 & 117294 \\
\hhline{|=|=|=|=|=|=|=|=|=|}
Speedup total time & 1.00 & 0.83 & 1.39 & 0.66 & 0.94 & 2.59 & 1.32 & 2.16 \\
Speedup linear solver time & 1.00 & 0.81 & 1.56 & 0.64 & 0.97 & 3.56 & 1.41 & 2.72 \\
\hline
\end{tabular}
\end{center}
\end{table*}

\begin{table*}[!h]
\caption{Breakdown of linear solve time (in seconds) spent running the solvers in flow with reduction=1e-2 (default)}
\begin{center}
\label{tab:flow_results_accelerator_breakdown}
\begin{tabular}{|c|l||c||c||c|c|c||c|c|c|}
\hline
\multirow{2}{*}{Device} & \multirow{2}{*}{Step (accumulated time)} &\multirow{2}{*}{CPU} & \multirow{2}{*}{FPGA} & \multicolumn{3}{c||}{GPU$^1$} & \multicolumn{3}{c|}{GPU$^2$} \\
\cline{5-10}
&&&& cusparse & OpenCL LS & OpenCL GC & cusparse & OpenCL LS & OpenCL GC \\
\hline
\multirow{7}{*}{Host} & Reorder vectors & - & 1.10 & - & 1.08 & 1.14 & - & 1.14 & 1.30 \\
\cline{3-10}
& Create preconditioner & 38.5 & 128.75 & 22.27 & 60.09 & 95.35 & 5.76 & 58.00 & 81.00 \\
& \hspace{8pt}BILU0 reorder matrix  & - & - & - & 20.47 & 27.72 & - & 21.30 & 23.60 \\
& \hspace{8pt}BILU0 decomposition   & - & - & - & 31.42 & 57.34 & - & 30.20 & 49.60 \\
& \hspace{8pt}BILU0 copy to GPU     & - & - & - & 5.04 & 5.97 & - & 3.79 & 4.63 \\
\cline{3-10}
& Memory setup & - & 10.61 & - & - & - & - & - & - \\
\cline{3-10}
& CPU to accelerator transfer  & - & 65.95 & 4.70 & 5.10 & 6.18 & 3.83 & 4.46 & 5.59 \\
\hline
\multirow{4}{*}{Accelerator} & Total solve & 274.1 & 304.90 & 175.10 & 448.50 & 324.10 & 71.80 & 197.30 & 88.10 \\
& \hspace{8pt}ILU apply             & 112.6 & - & 134.50 & 395.10 & 68.00 & 63.70 & 182.30 & 62.30 \\
& \hspace{8pt}SpMV                  & 127.3 & - & 28.40 & 37.30 & 25.50 & 2.60 & 4.40 & 7.70 \\
& \hspace{8pt}Rest                  & 28.9 & - & 8.50 & 12.70 & 223.80 & 4.00 & 8.30 & 14.20 \\\hline
Host & Accelerator to CPU transfer  & - & 0.76 & 0.27 & 0.35 & 0.16 & 0.26 & 0.34 & 0.16 \\
\hline
\end{tabular}
\end{center}
\end{table*}

To gain a better understanding of which parts of running the FPGA solver are causing it to be slower than the software version, we timed the different functions applied by this solver during a flow run, the results of which are shown in Table \ref{tab:flow_results_all}. During the running of the flow simulator, the sparsity pattern of the matrix that needs to be solved does not change. This makes it possible for functions that depend only on the sparsity pattern, like the finding of the re-ordering and partitioning pattern, to be done only once during initialization. From Table \ref{tab:flow_results_accelerator_breakdown}, we observe that the majority of solver time is actually spent running the FPGA solver, but a significant portion of the run time is also taken up by the creation of the preconditioner, and the data transfers between the host and FPGA.  

\section{Conclusion}
\label{sec:conclusion}

In this paper, we have evaluated the potential of using \acrshort{fpga}s in HPC, which is a highly relevant topic because of the rapid advances in reconfigurable hardware. To perform this study, we began by proposing a novel CSR based encoding to optimize a new SpMV kernel on \acrshort{fpga}, which can be easily integrated with an ILU0 preconditioner. We subsequently developed a hardware model to predict and guide the design of the ILU0 preconditioned BiCGstab solver, which helped us to understand the trade-offs between area and performance when scaling the resources. Next, we implemented the ILU0-BiCGStab preconditioned solver targeted at an HBM-enabled \acrshort{fpga} using a mixed programming model of HLS and RTL to maximize the performance of the double precision kernel. To validate our work, we integrated the complete solver in the OPM reservoir simulator, both for \acrshort{fpga} and GPU. Finally, we provided extensive evaluation results for both the standalone SpMV and solver kernels as well as the complete reservoir simulation execution on a real world use case running on three different platforms, CPU, \acrshort{fpga}, and \acrshort{gpu}.

We find that the \acrshort{fpga} is on par with the CPU execution and 3x slower than the \acrshort{gpu} implementation when comparing only the kernel executions. Although the obtained results show that an \acrshort{fpga} is not yet competitive with the \acrshort{gpu}, we believe that with a better on-chip support for double precision, increased number of computation units and memory ports, increased internal parallelism, as well as other optimizations on the host software side (i.e. reduction of preconditioner computation time, increased efficiency of the host-FPGA data exchange and task execution control), the \acrshort{fpga} can become a better alternative to speed-up scientific computing. This has been illustrated by scaling our hardware performance model beyond what it was possible to achieve in this work due to the encountered hardware limitations. In the future, we will work to eliminate these restrictions and provide a more efficient implementation. An alternative research could focus on developing other preconditioned solvers that might be more suitable for an \acrshort{fpga} implementation.

To encourage research into this direction, the source codes for the RTL kernels and the integration with \gls{opm} \textit{Flow} are made available in the \gls{opm} git repository \cite{opmrepo}. A \emph{pull request} was created (\#2998), which can be used to reproduce the results of this paper. Please note that since the \emph{pull request} was created against the most current version of the code at the time of writing (December 2020), the performance results obtained with that code may not be directly comparable with the results reported in this paper, which were gathered with a previous release of \gls{opm}'s \textit{Flow} as stated in section \ref{sec:flow_results}.\\
\\
%\input{sources/09_acknowledgements.tex}




%\bibliographystyle{IEEEtran}
\bibliography{main}{}
\bibliographystyle{IEEEtran}


% \end{thebibliography}

\end{document}